\documentclass[12pt]{article}
\usepackage{amsthm}
\usepackage{eurosym}
\usepackage{amsfonts}
\usepackage{amssymb,amsmath}
\usepackage[dvips]{color}
\usepackage{amscd}
\usepackage{tikz}
\usepackage{mathtools}
\usepackage{graphicx}
\usepackage{caption}
\usepackage{subcaption}
\usepackage[all,cmtip]{xy}

\newcommand{\xdownarrow}[1]{ {\left\downarrow\vbox to #1{}\right.\kern-\nulldelimiterspace} }
\usetikzlibrary{decorations.pathreplacing}

\usetikzlibrary{patterns}

\def\and{\mathrm{and}}

\newtheorem{lemm}{Lemma}

\newtheorem{remark}{Remark}
\newcommand{\ee}{\end{equation}}
\newcommand{\bea}{\begin{eqnarray}}
\newcommand{\eea}{\end{eqnarray}}
\newcommand{\beas}{\begin{eqnarray*}}
\newcommand{\eeas}{\end{eqnarray*}}
\newcommand{\ba}{\begin{array}}
\newcommand{\ea}{\end{array}}

\newcommand{\nbox}{{\,\lower0.9pt\vbox{\hrule \hbox{\vrule height 0.2 cm \hskip 0.19 cm \vrule height 0.2 cm}\hrule}\,}}

\def\href#1#2{#2}
\newcommand{\Cbb}{\mathbb{C}}
\theoremstyle{plain}

\newcommand{\Dc}{\mathcal{D}}

\newcommand{\Mfr}{\mathfrak{M}}

\newcommand{\Scl}{\mathcal{S}}

\newcommand{\Mcl}{\mathcal{M}}

\begin{document}

\begin{titlepage}
\hfill
\vbox{
    \halign{#\hfil         \cr
           } % end of \halign
      }  % end of \vbox

\hbox to \hsize{{}\hss \vtop{ \hbox{}

}}

%\begin{flushright}
%
%\end{flushright}

\vspace*{20mm}

\begin{center}

{\large \textbf{Boundary contributions to three loop superstring amplitudes}}

{\normalsize \vspace{10mm} }

{\normalsize {Kowshik Bettadapura${}^{1}$, Hai Lin${}^{1,2}$}  }

{\normalsize \vspace{10mm} }

{\small \emph{${}^1$\textit{Yau Mathematical Sciences Center, Tsinghua University,
Beijing 100084, China
}} }

{\normalsize \vspace{0.2cm} }

{\small \emph{$^2$\textit{Department of Mathematical Sciences, Tsinghua University,
Beijing 100084, China
\\
}} }

{\normalsize \vspace{0.4cm} }

\end{center}

\begin{abstract}

In type II superstring theory, the vacuum amplitude at a given loop order $g$ can receive contributions from the boundary of the compactified, genus $g$ supermoduli space of curves $\overline{\mathfrak M}_g$. These contributions capture the long distance or infrared behaviour of the amplitude. The boundary parametrises degenerations of genus $g$ super Riemann surfaces. A holomorphic projection of the supermoduli space onto its reduced space would then provide a way to integrate the holomorphic, superstring measure and thereby give the superstring vacuum amplitude at $g$-loop order. However, such a projection does not generally exist over the bulk of the supermoduli spaces in higher genera. Nevertheless, certain boundary divisors in $\partial\overline{\mathfrak M}_g$ may holomorphically map onto a bosonic space upon composition with universal morphisms, thereby enabling an integration of the holomorphic, superstring measure here. Making use of ansatz factorisations of the superstring measure near the boundary, our analysis shows that the boundary contributions to the three loop vacuum amplitude will vanish in closed oriented type II superstring theory with unbroken spacetime supersymmetry.

\end{abstract}

\end{titlepage}

\vskip 1cm

\section{Introduction}

\label{sec_Introduction} \renewcommand{\theequation}{1.\arabic{equation}} %
\setcounter{equation}{0}

A super Riemann surface of genus $g$ is a $(1|1)$-dimensional supermanifold
which is locally a superspace. Its reduced space is a genus $g$, Riemann
surface and its module of odd differentials defines a spin structure on the
surface. In contrast to punctures on Riemann surfaces however, there can be
two types of punctures on a super Riemann surface originating from external
states in superstring theory. These are Neveu-Schwarz (NS) type punctures
and the Ramond (R) type punctures.

\vspace{1pt}

The moduli space of super Riemann surfaces is referred to generally as \emph{%
supermoduli space}. It is a subjest of much importance in several areas of
mathematics and physics. It provides a unified framework in which to study
problems in algebraic geometry, the mathematics of supersymmetry and
superconformal quantum field theory. Methods from supergeometry and
superanalysis are essential for studying supersymmetric string theory from
the viewpoint of supermoduli space. For an early introduction to the
mathematics of supergeometry and superanalysis, see \cite{Berezin,Bartocci
etal,Voronov 01} and references therein.

\vspace{1pt}

In this paper we are interested in superstring amplitudes and the boundary
components of supermoduli spaces. More precisely, we aim to understand
contributions to the superstring amplitude at $g$ loop order arising from
the boundary of the compactified, genus $g$ supermoduli space $\overline{%
\mathfrak{M}}_{g}$. As an illustration of our ideas we specialise to genus $%
g=3$ and analyse the boundary contributions to the three loop vacuum
amplitude in Type II superstring theory.

\vspace{1pt}

More generally, the supermoduli space $\mathfrak{M}_{g,n,n^{\prime }}$ is
the moduli space of genus $g$ super Riemann surfaces with $n$ Neveu-Schwarz
punctures and $n^{\prime }$ Ramond punctures. In particular, the
dimensionality of the supermoduli space is increased by $(1|1)$ for each
Neveu-Schwarz puncture and by $(1|\frac{1}{2})$ for each Ramond puncture.
Two Ramond punctures contribute to one fermionic modulus. These moduli in
superstring perturbation theory play the role of the Schwinger parameters in
quantum field theory. In the context of a supersymmetric theory, fermionic
moduli play the role of supersymmetric partners to the bosonic moduli.

\vspace{1pt}

The Mumford isomorphisms in algebraic geometry are a collection of
isomorphisms between certain line bundles over the Deligne-Mumford
compactification of the moduli space of Riemann surfaces. In bosonic string
theory, defined by the path integral with the Polyakov action functional,
these isomorphisms can be used to construct the holomorphic string measure
\cite{Belavin:1986cy,Beilinson:1986zw}. Due to the existence of tachyons,
bosonic string theory is not entirely a physical theory, in contrast to its
supersymmetric extension---superstring theory. In superstring theory
however, the non-compact superstring configuration space is no longer the
moduli space of Riemann surfaces but rather its supersymmetric analogue, the
supermoduli space. Accordingly, the Mumford isomorphisms can be generalised
to isomorphisms between certain line bundles on supermoduli space \cite%
{Voronov:1988ms}, leading thereby to a construction of the holomorphic,
superstring measure \cite{Witten:2012bh}. For a construction of this measure
in the presence of NS- and R-punctures, see \cite{Witten:2013tpa} and \cite%
{Diroff:2018hua}.

\vspace{1pt}

The superstring measure on the supermoduli space can also be derived from
worldsheet superconformal field theory \cite%
{Witten:2012bh,DHoker:1988pdl,Giddings:1987im,Friedan:1985ey,Moore:1985uh,Verlinde:1987sd,AlvarezGaume:1988sj}%
. The integration of this measure over the supermoduli space then gives the
superstring amplitude. A chiral splitting procedure is important \cite%
{DHoker:1988pdl,DHoker:1989cxq,Verlinde:1987sd} in the computation of the
genus two superstring measure \cite{DHoker:2001kkt}. In the chiral splitting
procedure, one introduces loop momenta to write conformal correlators via an
integral over loop momenta whose integrand is the product of the left and
right chiral conformal blocks. Each chiral block is analytic in the moduli
of the surface as well as in the inserted vertex points.

\vspace{1pt}\vspace{1pt}

Integration of the genus $g$, bosonic string measure over the
Deligne-Mumford compactifcation of the bosonic moduli space, i.e. the moduli
space of Riemann surfaces, gives the $g$ loop bosonic string amplitude.
Similarly and as mentioned above, integration of the genus $g$, superstring
measure gives a contribution to the superstring amplitude at $g$ loop order.
The space over which one integrates is the analogue of the compactification
of bosonic moduli space, being a compactification of supermoduli space. Such
a compactification was sketched by Deligne in a letter to Manin circa 1987,
and described recently and in more detail by Witten in \cite{Witten:2012ga}
and Donagi and Witten in \cite{Donagi:2013dua}. Expressions for the
superstring amplitude in genus $g=0,1$ were obtained by Green and Schwarz
\cite{Green:1981yb}, and in $g=2$ by D'Hoker and Phong \cite{DHoker:2001kkt}%
. However, as observed by Witten \cite{Witten:2015hwa}, these expressions
ought to receive potentially non-vanishing contributions from the boundary
divisors in the compactification of supermoduli space. At two loop order,
Witten nevertheless notes that these boundary contributions will vanish. In
this paper we note that the boundary contributions of vacuum amplitude at
three loop order will also vanish.

\vspace{1pt}

From more mathematical perspectives, one of the main results by Donagi and
Witten \cite{Donagi:2013dua,Donagi:2014hza} concerns the question of
holomorphically projecting the genus $g$ supermoduli space $\mathfrak{M}_{g}$
onto its reduced space. A holomorphic projection $\pi :\mathfrak{M}%
_{g}\rightarrow \mathcal{S}\mathcal{M}_{g}$ would allow for computing the
superstring amplitude by firstly integrating along the odd fibers as
stipulated by Berezin \cite{Berezin}, and then integrating over the spin
moduli space $\mathcal{S}\mathcal{M}_{g}$. Since the spin moduli space
discretely covers the moduli space of Riemann surfaces $\mathcal{M}_{g}$,
measures on $\mathcal{S}\mathcal{M}_{g}$ can be reduced to measures on $%
\mathcal{M}_{g}$ by summing over the spin structures---a procedure known as
GSO projection. This method of reducing the superstring measure to a measure
on $\mathcal{M}_{g}$ was used by Green and Schwarz \cite{Green:1981yb} and
by D'Hoker and Phong \cite{DHoker:2001kkt} in their derivation of the
superstring amplitude to loop orders zero, one and two. Donagi and Witten
found however that $\mathfrak{M}_{g}$ cannot be projected onto $\mathcal{S}%
\mathcal{M}_{g}$ for any genus $g\geq 5$, thereby placing a mathematical
obstruction to applying known methods to calculate superstring amplitudes to
arbitrary loop order. Note, this is in contrast to bosonic string theory
where the bosonic string amplitudes are in principle known to any loop order.

\vspace{1pt}

Supermoduli space parametrises smooth, super Riemann surfaces. In its
compactification, boundary divisors parametrise super Riemann surfaces which
can degenerate into two kinds, separating degenerations and non-separating
degenerations. These boundary divisors can be identified with punctured
supermoduli spaces of generally lower genera through a process known as
clutching, as studied in a more classical setting in \cite{Knudsen}. The
boundary contributions to superstring amplitudes involve integrating the
superstring measure over these boundary divisors. In this paper we look at
these boundary contributions along boundary divisors in the genus $g=3$
supermoduli space. As observed earlier, there do not exist holomorphic
projections of supermoduli in genus $g\geq 5$. In genus $g=3$ there does
exist a projection, but it is not holomorphic---it is singular along the
hyperelliptic locus \cite{Witten:2015hwa}. As such, although one can use
super period matrix \cite{Witten:2015hwa,DHoker:2015gwa} as a basis, it is
unclear as to how to apply D'Hoker and Phong's integration procedure over
the genus $g=3$ supermoduli space. As mentioned above, we have largely
considered boundary components of the supermoduli space, which involve
supermoduli spaces of lower genera.

\vspace{1pt}

The organisation of this paper is as follows. In Section 2, we describe the
boundary components of supermoduli space. This section is divided into three
subsections which emphasise perspectives from algebraic geometry, from
gluing in local, geometric models, and from superstring worldsheet theory.
Section 3 is devoted to superstring measures and contributions to the
amplitude from the boundary of supermoduli space. This section is divided
into five subsections offering perspectives on superstring measures and
amplitudes from supergeometry and sheaf theory in algebraic geometry. We
analyse the genus three case as a detailed example. In Section 4 we discuss
our results and draw conclusions. In the interests of being self-contained,
we give a brief overview of the compactification of the moduli space of
Riemann surfaces with and without spin structures in Appendix A.

\vspace{1pt}

\section{Boundary of supermoduli spaces}

\label{sec_intro_} \renewcommand{\theequation}{2.\arabic{equation}}
\renewcommand{\thethm}{2.\arabic{thm}} \setcounter{thm}{0} %
\renewcommand{\theprop}{2.\arabic{prop}} \setcounter{prop}{0}

In this section we analyse the boundary of supermoduli space and related
aspects. The boundary parametrises super Riemann surfaces with prescribed
degenerations. That is, super Riemann surfaces which develop nodes of
Neveu-Schwarz (NS) or Ramond (R) type. As described in \cite{Witten:2012ga},
there are two distinct types of degenerations of a super Riemann surface,
referred to as separating and non-separating degenerations. The construction
of the boundary of supermoduli space parallels that of the boundary of the
Deligne-Mumford compactification of the moduli space of Riemann surfaces in
\cite{Knudsen} and \cite{Deligne Mumford}. In Section 2.1 we analyse the
boundary components of supermoduli space parametrising degenerating super
Riemann surfaces. We emphasise the role of clutching morphisms, among other
things. Section 2.2 reviews gluing formulae for these degenerating surfaces
at Neveu-Schwarz and Ramond nodes. In Section 2.3 we describe the relation
between the superstring amplitudes and the boundary of the supermoduli
spaces.

\subsection{Boundary components via clutching}

Like the objects it parametrises, the moduli space of genus $g$ super
Riemann surfaces $\mathfrak{M}_{g}$ is a superspace. Its reduced space $%
\mathcal{S}\mathcal{M}_{g}$ parametrises Riemann surfaces with spin
structure. As such, its compactification might resemble that of the spin
moduli space described by Cornalba \cite{Cornalba} and which in turn
resembles the Deligne-Mumford compactification \cite{Deligne Mumford}. And
indeed, the compactification of ${\mathfrak{M}}_{g}$ is similar to that of $%
\mathcal{M}_{g}$. To be self-contained, we give a brief overview of the
compactifications of the ordinary bosonic moduli space $\mathcal{M}_{g}$ and
spin moduli space $\mathcal{S}\mathcal{M}_{g}$ in Appendix A.

Denote by $\mathfrak{M}_{g,n,n^{\prime }}$ the moduli space of super Riemann
surfaces of genus $g$ with $n$-many NS-punctures and $n^{\prime }$-many
R-punctures. We use a prime in denoting the R puncture number so as to
distinguish them from the NS puncture number. The reduced space of $%
\mathfrak{M}_{g,n,n^{\prime }}$ is $\mathcal{S}\mathcal{M}_{g,n+n^{\prime }}$%
. If $n=n^{\prime }=0$, we will just write $\mathfrak{M}_{g}$ rather than $%
\mathfrak{M}_{g,0,0^{\prime }}$ for brevity. Similarly, if either $n=0$ or $%
n^{\prime }=0$, the respective subscript will be omitted. As with the case
of spin moduli space, the parity of spin structures induces a decomposition
of the supermoduli space: $\mathfrak{M}_{g,n,n^{\prime }}=\mathfrak{M}%
_{g,n,n^{\prime };+}\cup \mathfrak{M}_{g,n,n^{\prime };-}$, where $+$ or $-$%
~denotes even or odd for the parity of the spin structure. It is $\mathfrak{M%
}_{g,n,n^{\prime };+}$ which is of primary interest in superstring theory.
But as illustrated in the boundary components of spin moduli space in
Appendix A, the boundary of $\partial \overline{\mathfrak{M}}_{g,n,n^{\prime
};+}$ will inevitably involve odd spin structures. Now in contrast to
Neveu-Schwarz punctures, there must always be an even number of Ramond
punctures, i.e., $n^{\prime }$ is an even integer. In genus $g\geq 2$,
appropriately applying the Riemann-Roch theorem reveals the dimension:
\begin{equation}
\dim \mathfrak{M}_{g,n,n^{\prime }}=3g-3+n+n^{\prime }~\big|~2g-2+n+\frac{1}{%
2}n^{\prime }.
\end{equation}%
In analogy with stable curves, compactifying $\mathfrak{M}_{g,n,n^{\prime }}$
involves allowing super Riemann surfaces to develop nodes of Neveu-Schwarz
or Ramond type. The boundary components $\partial \overline{\mathfrak{M}}%
_{g,n,n^{\prime }}\subset \overline{\mathfrak{M}}_{g,n,n^{\prime }}$ ought
then be similar to those of $\overline{\mathcal{M}}_{g,n}$ from (\ref%
{clutching_bosonic}). This is indeed the case along the Neveu-Schwarz nodes.
As such, components of $\partial \overline{\mathfrak{M}}_{g,n,n^{\prime }}$
are given by the analogue of $\alpha $- and $\beta $-type clutchings in (\ref%
{clutching_bosonic}), (\ref{alpha_spineven}) and (\ref{beta_spinodd}) for
super Riemann surfaces along NS nodes. This leads now to the following
clutching morphisms for the boundary component of the even part of
supermoduli space:%
\begin{equation}
\overline{\mathfrak{M}}_{g_{1},n_{1}+1,n_{1}^{\prime };\pm }\times \overline{%
\mathfrak{M}}_{g_{2},n_{2}+1,n_{2}^{\prime };\pm }\overset{\alpha ^{(\pm
,\pm )}}{\longrightarrow }\overline{\mathfrak{M}}_{g,n,n^{\prime };+}~~~%
\mbox{and}~~~\overline{\mathfrak{M}}_{g-1,n+2,n^{\prime };+}\overset{\beta
^{+}}{\longrightarrow }\overline{\mathfrak{M}}_{g,n,n^{\prime };+}
\label{clutching morphism}
\end{equation}%
where $g_{1}+g_{2}=g$,$~n_{1}+n_{2}=n$ and $n_{1}^{\prime }+n_{2}^{\prime
}=n^{\prime }$. There are similar clutching morphisms to (\ref{clutching
morphism}) describing the boundary of the odd part of supermoduli space.
Note that accordingly we have the inclusions,
\begin{equation}
\mathrm{im}~\alpha \subset \partial \overline{\mathfrak{M}}_{g,n,n^{\prime
}}\subset \overline{\mathfrak{M}}_{g,n,n^{\prime }},~~~\ ~~~\mathrm{im}%
~\beta \subset \partial \overline{\mathfrak{M}}_{g,n,n^{\prime }}\subset
\overline{\mathfrak{M}}_{g,n,n^{\prime }},  \label{inclusions}
\end{equation}%
where $\alpha $ and $\beta $ are as in (\ref{clutching morphism}). We
sometimes denote $\alpha =\alpha _{g_{1},g_{2}}$ and$~\beta =\beta _{g-1}$
with the subscript indicating the genus; and sometimes we omit the subscript
for brevity. If $\mathcal{D}_{NS;sep.}$ and $\mathcal{D}_{NS;nsep.}$ in $%
\partial \overline{\mathfrak{M}}_{g,n,n^{\prime }}$ denote the divisors
parametrising separating and non-separating degenerations along NS-nodes,
then:
\begin{equation}
\mathcal{D}_{NS;sep.;\pm }\cong \overline{\mathfrak{M}}%
_{g_{1},n_{1}+1,n_{1}^{\prime };\pm }\times \overline{\mathfrak{M}}%
_{g_{2},n_{2}+1,n_{2}^{\prime };\pm }~~~~\mathrm{and}~~~~\mathcal{D}%
_{NS;nsep.;+}\cong \overline{\mathfrak{M}}_{g-1,n+2,n^{\prime };+}.
\label{boundary_03}
\end{equation}%
Note $\mathcal{D}_{NS;sep.}=\mathcal{D}_{NS;sep.;+}\cup \mathcal{D}%
_{NS;sep.;-}$ and similarly for $\mathcal{D}_{NS;nsep.}$. These divisors
form the boundary components of $\overline{\mathfrak{M}}_{g,n.n^{\prime }}$
corresponding to degenerations along NS nodes. The divisors parametrising
degenerations along Ramond nodes cannot be so elegantly described however.
As discussed by Witten \cite{Witten:2012ga}, the divisors $\mathcal{D}%
_{R;sep.}$ and $\mathcal{D}_{R;nsep.}$ in $\partial \overline{\mathfrak{M}}%
_{g,n,n^{\prime }}$ parametrising separating and non-separating
degenerations along Ramond nodes are fibered over the expected boundary
components with fermionic fiber. More generally, if $\mathfrak{X}$ is a
supermanifold with boundary $\partial \mathfrak{X}$, the boundary $\partial
\mathfrak{X}\subset \mathfrak{X}$ will have codimension-$(1|0)$. Dimension
counting now reveals, along a separating Ramond degeneration:
\begin{align}
\dim \mathfrak{M}_{g,n,n^{\prime }}-(\dim \mathfrak{M}_{g_{1},n_{1},n_{1}^{%
\prime }}& +\dim \mathfrak{M}_{g_{2},n_{2},n_{2}^{\prime }})  \notag \\
& =n^{\prime }-(n_{1}^{\prime }+n_{2}^{\prime })+3~\big|~\frac{1}{2}%
(n^{\prime }-(n_{1}^{\prime }+n_{2}^{\prime }))+2  \label{dimension_03}
\end{align}%
where we have taken $g=g_{1}+g_{2}$ and $n=n_{1}+n_{2}$. Since the number of
Ramond punctures must be even, and since $n^{\prime }<n_{1}^{\prime
}+n_{2}^{\prime }$, the dimension formula in (\ref{dimension_03}) only makes
sense when $n^{\prime }+2=n_{1}^{\prime }+n_{2}^{\prime }$. Evidently, the
divisor $\mathcal{D}_{R;sep.}\subset \partial \overline{\mathfrak{M}}%
_{g,n,n^{\prime }}$ parametrising separating degeneration along Ramond
punctures is a $(0|1)$-dimensional fibration over $\overline{\mathfrak{M}}%
_{g_{1},n_{1},n_{1}^{\prime }}\times \overline{\mathfrak{M}}%
_{g_{2},n_{2},n_{2}^{\prime }}$. A similar analysis in the non-separating
case $\mathcal{D}_{R;nsep.}\subset \partial \overline{\mathfrak{M}}%
_{g,n,n^{\prime }}$ reveals it will fiber over $\overline{\mathfrak{M}}%
_{g-1,n,n^{\prime }+2}$ with $(0|1)$-dimensional fibers. And so, the
boundary components $\mathcal{D}_{R;sep.},\mathcal{D}_{R;nsep.}\subset
\partial \overline{\mathfrak{M}}_{g,n.n^{\prime }}$ parametrising separating
and non-separating Ramond degenerations can be realised as fibrations:
\begin{equation}
\xymatrix{ \Cbb^{0|1}\ar[r]^\subset & \Dc_{R; sep.;\pm} \ar[d]^{\pi_{R;
sep.;\pm}} \\ &\overline{\Mfr}_{g_1, n_1, n_1^\prime;\pm}\times
\overline{\Mfr}_{g_2, n_2, n_2^\prime;\pm} }\quad \mathrm{and}\quad \quad %
\xymatrix{ \Cbb^{0|1}\ar[r]^\subset & \Dc_{R; nsep.;+} \ar[d]^{\pi_{R;
nsep.;+}} \\ & \overline{\Mfr}_{g-1, n, n^\prime+2;+} }  \label{boundary_04}
\end{equation}%
where $g_{1}+g_{2}=g$, $n_{1}+n_{2}=n$ and $n_{1}^{\prime }+n_{2}^{\prime
}=n^{\prime }+2$. Note the contrast with NS-case in (\ref{boundary_03}). As
we will see in Section 2.2, this contrast will be apparent in the gluing
laws along the nodes in the following sense: there are no free, odd
parameters in the gluing laws along NS nodes. As in the Neveu-Schwarz case,
the divisors parametising Ramond degenerations decompose according to the
parity of spin structures, i.e., $\mathcal{D}_{R;sep.}=\mathcal{D}%
_{R;sep.;+}\cup \mathcal{D}_{R;sep.;-}$, and similarly for $\mathcal{D}%
_{R;nsep.}$

We can now write the boundary of supermoduli space as follows:
\begin{equation}
\partial \overline{\mathfrak{M}}_{g,n,n^{\prime }}=\partial \overline{%
\mathfrak{M}}_{g,n,n^{\prime };NS}\cup \partial \overline{\mathfrak{M}}%
_{g,n,n^{\prime };R}.  \label{decomposition}
\end{equation}%
The compactified supermoduli space is now the union of the bulk with the
boundary,%
\begin{equation}
\overline{\mathfrak{M}}_{g,n,n^{\prime }}=\mathfrak{M}_{g,n,n^{\prime }}\cup
\partial \overline{\mathfrak{M}}_{g,n,n^{\prime }}.
\end{equation}

Another description of the boundary is as a union over irreducible
components $\partial \overline{\mathfrak{M}}_{g,n,n^{\prime
}}=\bigcup_{j}\Delta _{j}$, where $\Delta _{j}$ denotes an irreducible
component. Then to the embedding $\Delta _{j}\subset \overline{\mathfrak{M}}%
_{g,n,n^{\prime }}$ there is a bundle of normal sections to the embedding
over $\Delta _{j}$, denoted $\mathcal{N}_{\Delta _{j}}$. Its sheaf of
sections, denoted $\hat{\mathcal{N}}_{\Delta _{j}}$, fits into a short exact
sequence
\begin{equation}
0\rightarrow T_{\Delta _{j}}\rightarrow T_{\overline{\mathfrak{M}}%
_{g,n,n^{\prime }}}|_{\Delta _{j}}\rightarrow \hat{\mathcal{N}}_{\Delta
_{j}}\rightarrow 0,
\end{equation}%
or dually%
\begin{equation}
0\rightarrow \hat{\mathcal{N}}_{\Delta _{j}}^{\ast }\rightarrow T_{\overline{%
\mathfrak{M}}_{g,n,n^{\prime }}}^{\ast }|_{\Delta _{j}}\rightarrow T_{\Delta
_{j}}^{\ast }\rightarrow 0.
\end{equation}%
We denote by $\mathfrak{N}_{\Delta _{j}}$ the fiber of the normal bundle $%
\mathcal{N}_{\Delta _{j}}$ to the embedding $\Delta _{j}\subset \overline{%
\mathfrak{M}}_{g,n,n^{\prime }}$. This gives a fibration of spaces,
\begin{equation}
\mathfrak{N}_{\Delta _{j}}\rightarrow \mathcal{N}_{\Delta _{j}}\rightarrow
\Delta _{j}.  \label{normal bundle 03}
\end{equation}%
As mentioned earlier, the sheaf of sections of $\mathcal{N}_{\Delta _{j}}$
over $\Delta _{j}$ is $\hat{\mathcal{N}}_{\Delta _{j}}$.

As in the case of Riemann surfaces with marked points, we can form a
forgetful map on super Riemann surfaces,
\begin{equation}
\overline{\mathfrak{M}}_{g,n+1,n^{\prime }}\overset{p}{\longrightarrow }%
\overline{\mathfrak{M}}_{g,n,n^{\prime }},  \label{forgetful_morphism_01}
\end{equation}%
given by forgetting the ($n+1$)-th NS puncture and stabilizing the resulting
punctured super Riemann surface. The pull-back along the forgetful map is
associated to the insertion of an vertex operator in superconformal field
theory. We also denote $p=p_{g}$ with the subscript meaning the genus.
Sometimes we omit the subscript for brevity. Composite forgetful maps gives
the morphism $\overline{\mathfrak{M}}_{g,n+2,n^{\prime }}\overset{p^{2}}{%
\longrightarrow }\overline{\mathfrak{M}}_{g,n,n^{\prime }}$.

The following are diagrams of morphisms relevant for subsequent
constructions in this paper:
\begin{equation}
\begin{array}{ccccc}
\overline{\mathfrak{M}}_{g_{1},n_{1}+1,n_{1}^{\prime }} & \times & \overline{%
\mathfrak{M}}_{g_{2},n_{2}+1,n_{2}^{\prime }} & \overset{\alpha }{%
\xrightarrow{\hspace*{1cm}}} & \overline{\mathfrak{M}}_{g,n,n^{\prime }} \\
{\Big\downarrow}%
\begin{array}{c}
_{{p}} \\
\end{array}
&  & {\Big\downarrow}%
\begin{array}{c}
_{{p}} \\
\end{array}
&  &  \\
\overline{\mathfrak{M}}_{g_{1},n_{1},n_{1}^{\prime }} & \times & \overline{%
\mathfrak{M}}_{g_{2},n_{2},n_{2}^{\prime }} &  &
\end{array}%
\end{equation}%
and%
\begin{equation}
\begin{array}{ccc}
\overline{\mathfrak{M}}_{g-1,n+2,n^{\prime }} & \overset{\alpha }{%
\xrightarrow{\hspace*{1cm}}} & \overline{\mathfrak{M}}_{g,n,n^{\prime }} \\
{\Big\downarrow}%
\begin{array}{c}
_{{p}^{2}} \\
\end{array}
&  &  \\
\overline{\mathfrak{M}}_{g-1,n,n^{\prime }} &  &
\end{array}%
\end{equation}

The clutching morphisms are defined on boundary components of supermoduli
spaces which parametrise degenerating super Riemann surfaces. These
degenerations can be realised in a geometric model involving gluing maps in
Section 2.2. They are important for Section 3.3.

In this section we have described the compactification of supermoduli space.
In Section 3.2 we consider a `partial compactification' of supermoduli
space. This is in order to ensure that a certain, holomorphic map defined on
the bulk will remain holomorphic when extended to the (partial)
compactification.

\vspace{1pt}

\subsection{Degenerations and gluing maps in a geometric model}

The degeneration of super Riemann surfaces parametrised by the boundary of
supermoduli spaces can be realised in a geometric model via gluing. These
models are necessarily local however, in the sense that the gluing map is a
local map. It is only defined near the node along which the super Riemann
surface degenerates.

As described in \cite{Witten:2012bh}, we can glue super Riemann surfaces $%
\mathcal{S}_{\ell }$ and $\mathcal{S}_{r}$ with local coordinates $x|\theta $
and $y|\psi $, respectively. We glue them so that the gluing happens at the
points $a|\alpha \in \mathcal{S}_{\ell }$ and $b|\beta \in \mathcal{S}_{r}$,%
\begin{eqnarray}
(x-a-\alpha \theta )(y-b-\beta \psi ) &=&-\varepsilon ^{2}~=~q_{NS},  \notag
\\
(y-b-\beta \psi )(\theta -\alpha ) &=&\varepsilon (\psi -\beta ),  \notag \\
(x-a-\alpha \theta )(\psi -\beta ) &=&-\varepsilon (\theta -\alpha ),  \notag
\\
(\theta -\alpha )(\psi -\beta ) &=&0.  \label{gluing 01}
\end{eqnarray}%
In the language of theoretical physics, the superconformal structures are
defined by the vector fields $D_{\theta }=\partial _{\theta }+\theta
\partial _{x}$ and $D_{\psi }=\partial _{\psi }+\psi \partial _{y}$ on the
surfaces $\mathcal{S}_{\ell }$ and $\mathcal{S}_{r}$ respectively. The
gluing formula maps the superconformal coordinates of the left component $%
\mathcal{S}_{\ell }$ to the superconformal coordinates of the right
component $\mathcal{S}_{r}$. The above (\ref{gluing 01}) is gluing along the
NS punctures and then smoothed out. The $\alpha ,\beta $ are fermionic
parameters.

The Ramond punctures define divisors on the super Riemann surface \cite%
{Witten:2012bh}, of dimension $0|1$. We can also glue along these divisors
in a manner analogous to but more subtly than in (\ref{gluing 01}). In the
case of a separating Ramond degeneration, the Ramond punctures in $\mathcal{S%
}_{\ell }$ and $\mathcal{S}_{r}$ are glued and smoothed out. In local
coordinates $x|\theta $ on $\mathcal{S}_{l}$ and $y|\psi $ on $\mathcal{S}%
_{r}$ the superconformal structures are defined by the vector fields $%
\partial _{\theta }+x\theta \partial _{x}$ and $\partial _{\psi }+y\psi
\partial _{y}$ respectively. The divisors corresponding to the Ramond
punctures are defined respectively by $x=0$ and $y=0$; and the gluing is
\cite{Witten:2012bh}%
\begin{eqnarray}
xy &=&q_{R},~~  \notag \\
\theta &=&\zeta \pm \sqrt{-1}\psi .  \label{gluing 02}
\end{eqnarray}%
where $q_{R}$ is a bosonic gluing parameter. The free fermionic parameter $%
\zeta $ parametrizes the fiber of a component of the fibration $\mathcal{D}%
_{sep.;R}$, c.f., (\ref{boundary_04}). Evidently, each Ramond divisor has
fermionic fiber $\mathbb{C}^{0|1}$.

\vspace{1pt}

\subsection{Amplitudes from boundary contributions}

We would like to calculate amplitudes for the scattering of $n+n^{\prime }$
superstring states. These can be calculated by the correlation functions $%
\langle V_{1}\dots V_{n+n^{\prime }}\rangle ~$of vertex operators $%
V_{1},\dots ,V_{n+n^{\prime }}$ in worldsheet superconformal field theory.
We use the RNS formalism with manifest worldsheet supersymmetry to quantise
the theory. The superstring worldsheet is manifestly a super Riemann surface
and so the amplitude can also be calculated by integrating over supermoduli
space. To illustrate, consider a form $\hat{F}_{V_{1},\dots ,V_{n+n^{\prime
}}}\ $on the supermoduli space
\begin{equation}
\hat{F}_{V_{1},\dots ,V_{n+n^{\prime }}}=\int \mathcal{D}(X,B,C,\tilde{B},%
\tilde{C})\exp (-\hat{I})\prod_{i=1}^{n+n^{\prime }}V_{i}(p_{i}),
\end{equation}%
where $X$ denotes worldsheet matter fields which are also the spacetime
coordinates, $C,\tilde{C}$ denote worldsheet ghost fields, $B,\tilde{B}~$%
denote worldsheet antighost fields, $p_{i}$ denote points, and $\hat{I}$ is
the action of the worldsheet theory after gauge-fixing. In the simplest
formalism we use NS vertex operators of picture number $-1$, and Ramond
vertex operators of picture number $-1/2$.

The genus $g$ total contribution to the scattering amplitude is $\hat{%
\mathcal{A}}=\langle V_{1}\dots V_{n+n^{\prime }}\rangle $. We consider $%
\hat{\mathcal{A}}$ to be the summation of contributions from the bulk and
from the boundary of supermoduli space $\hat{\mathcal{A}}_{\mathrm{bulk}}=%
\hat{\mathcal{A}}-\mathcal{A}$ and $\hat{\mathcal{A}}_{\mathrm{bndy}}=%
\mathcal{A}~$respectively. The form restricted to the boundary is%
\begin{equation}
F_{V_{1},\dots ,V_{n+n^{\prime }}}=\hat{F}_{V_{1},\dots ,V_{n+n^{\prime }}}{%
\big |}_{\partial \overline{\mathfrak{M}}_{g,n,n^{\prime }}}.
\end{equation}%
The contribution to the amplitude, from the boundary of the supermoduli
space, where the super Riemann surfaces degenerate, is given by%
\begin{equation}
\mathcal{A}=\int_{\partial \overline{\mathfrak{M}}_{g,n,n^{\prime }}}\int_{%
\mathfrak{N}}F_{V_{1},\dots ,V_{n+n^{\prime }}}.  \label{amplitude_boundary}
\end{equation}%
Here $\mathfrak{N}$ is the fiber of the normal bundle to the boundary
divisor, discussed in (\ref{normal bundle 03}).

\section{Boundary contributions and amplitudes}

\renewcommand{\theequation}{3.\arabic{equation}} \setcounter{equation}{0} %
\renewcommand{\thethm}{3.\arabic{thm}} \setcounter{thm}{0} %
\renewcommand{\theprop}{3.\arabic{prop}} \setcounter{prop}{0} %
\renewcommand{\thelemm}{3.\arabic{lemm}} \setcounter{lemm}{0} %
\renewcommand{\theremark}{3.\arabic{remark}} \setcounter{remark}{0}

In this section we analyse contributions to amplitudes from the boundary of
supermoduli spaces. We begin in Section 3.1 by presenting an overview of the
Berezinian of bundles over supermoduli space. We present Lemma 3.1
concerning a particular class of Berezinians. These are relevant in the
formulation of the super Mumford isomorphisms and forms which define the
superstring measures. In Section 3.2 we investigate contributions from the
boundary of supermoduli space for a general genus. This involves firstly
constructing a holomorphic map from a partial compactification of
supermoduli space in genus two; and secondly, using this map as a building
block to extract the aforementioned boundary contributions. In Section 3.3
we use the geometric model of gluing maps to illustrate the pole behaviour
of the super Mumford form near the boundary divisors. In Section 3.4 we
overview the notion of integration on supermanifolds more generally, with
the aim to apply these notions to compute superstring amplitudes through
integration over supermoduli spaces. We present a useful integration formula
(\ref{integral 05}) which will be used. Section 3.5 serves as an example of
these formalisms for the three loop vacuum amplitude.

\subsection{Super Mumford isomorphisms and forms}

A supermanifold $\mathfrak{X}$ is a space which is modelled on the data of a
manifold $X$ and a vector bundle $T_{X,-}^{\ast }$ on $X$, thought of as the
module of `odd' differentials or `fermionic parameters'. The subscript $+$
or $-$ denote even or odd respectively, in this section. We can also think
of $T_{X,-}^{\ast }$ as a locally free sheaf on $X$. The space $X$ is itself
referred to as the reduced space of $\mathfrak{X}$. The prototype
supermanifold associated to $(X,T_{X,-}^{\ast })$ is the split model, which
is the locally ringed space $(X,\wedge ^{\bullet }T_{X,-}^{\ast })$. Its
dimension is defined by a pair $(p|q)$ where $p=\dim X$ and $q=\mathrm{rank}%
~T_{X,-}^{\ast }$. More generally, a supermanifold \emph{modelled} on $%
(X,T_{X,-}^{\ast })$ is a locally ringed space $\mathfrak{X}=(X,\mathcal{O}_{%
\mathfrak{X}})$ where $\mathcal{O}_{\mathfrak{X}}$ is a sheaf of (local)
supercommutative algebras on $X$, referred to as the structure sheaf. It is,
additionally, required to be locally isomorphic to $\wedge ^{\bullet
}T_{X,-}^{\ast }$ as sheaves over $X$. The dimension of $\mathfrak{X}$
coincides with the dimension of the split model. If $\mathfrak{X}$ is
isomorphic to the split model, it is split; otherwise, it is non-split. A
more relevant and weaker condition for the purposes of physics is a
(holomorphic) projection or fibration of $\mathfrak{X}$ over its reduced
space $X$. As we will discuss in sections to follow, the existence of a
projection allows for the reduction of measures on superspace to measures on
the reduced space where classical methods of integration can be applied.
Note, if $\mathfrak{X}$ is split, then it is projected; but not necessarily
conversely.

Since the structure sheaf $\mathcal{O}_{\mathfrak{X}}$ of a supermanifold $%
\mathfrak{X}=(X,\mathcal{O}_{\mathfrak{X}})$ is a sheaf of supercommutative
algebras, it is globally $\mathbb{Z}_{2}$-graded. Hence we can write $%
\mathcal{O}_{\mathfrak{X}}=\mathcal{O}_{\mathfrak{X},+}\oplus \mathcal{O}_{%
\mathfrak{X},-}$ as $\mathcal{O}_{\mathfrak{X},+}$-modules. The tangent
sheaf can be graded compatibly and so we have%
\begin{equation}
T_{\mathfrak{X}}=T_{\mathfrak{X},+}\oplus T_{\mathfrak{X},-}.
\label{tangent bundle 01}
\end{equation}%
If $\mathcal{J}=\mathcal{O}_{\mathfrak{X},-}\cdot \mathcal{O}_{\mathfrak{X}%
}\subset \mathcal{O}_{\mathfrak{X}}$ denotes the fermionic ideal, then the
structure sheaf of the reduced space $X$ is $\mathcal{O}_{X}=\mathcal{O}_{%
\mathfrak{X}}/\mathcal{J}$.~If $\mathfrak{X}$ is modelled on $%
(X,T_{X,-}^{\ast })$ then $T_{\mathfrak{X},\pm }\cong T_{X,\pm }~\mathrm{mod}%
~\mathcal{J}$, where $T_{X,+}=T_{X}$ is the tangent sheaf of $X$ and $%
T_{X,-} $ are the `odd' tangent vectors. Hence, if $\mathfrak{X}$ is a
supermanifold modelled on $(X,T_{X,-}^{\ast })$ with tangent sheaf as in (%
\ref{tangent bundle 01}), its Berezinian will then be given by
\begin{equation}
\mathrm{Ber}~\mathfrak{X}=\mathrm{Ber}~T_{\mathfrak{X}}^{\ast }~\cong \frac{%
\omega _{X}}{\det T_{X,-}^{\ast }}=\omega _{X}\otimes \det T_{X,-}~~~~~%
\mathrm{mod}~\mathcal{J}  \label{berezinian 01}
\end{equation}%
where $\omega _{X}=\det T_{X,+}^{\ast }$ is the canonical bundle of $X$.
Here, forming the denominator means tensoring by its dual.

A super Riemann surface of genus $g$ is a $(1|1)$-dimensional supermanifold $%
\mathcal{S}$ modelled on a Riemann surface $C$ and a spin structure $%
T_{C,-}^{\ast }$ as its module of odd differentials. Spin structures are
also referred to as theta characteristics \cite{Cornalba} in the language of
algebraic geometry. Recall that a line bundle on a Riemann surface is an
ordinary spin structure if its quadratic tensor power is isomorphic to the
canonical bundle. That the module of odd differentials must be a spin
structure follows from a more general characterisation: a super Riemann
surface $\mathcal{S}$ is a supermanifold with $C$ as its reduced space and a
choice of nowhere integrable distribution $\mathcal{D}\subset T_{\mathcal{S}%
} $, where $\mathcal{D}$ is generated by the superconformal vector fields on
the super Riemann surface. In denoting the superconformal vector field by $%
D_{\theta }=\partial _{\theta }+\theta w(x)\partial _{x}$, see that $%
D_{\theta }^{2}=w(x)\partial _{x}$. In particular, the function $w(x)$ has
zeros along the Ramond punctures where $D_{\theta }^{2}=0$. We denote the
divisors for Neveu-Schwarz or Ramond punctures on the super Riemann surface
as $\mathcal{P}_{i}$ or $\mathcal{F}_{i}$ respectively in our convention,
and their sum by $\ \mathcal{P}=\sum_{i=1}^{n}\mathcal{P}_{i}$,$~\ \mathcal{F%
}=\sum_{i=1}^{n^{\prime }}\mathcal{F}_{i}$. On a super Riemann surface then
with $n$ Neveu-Schwarz punctures and $n^{\prime }$ Ramond punctures, the
distribution $\mathcal{D}$ sits in a short exact sequence
\begin{equation}
0\rightarrow \mathcal{D}\rightarrow T_{\mathcal{S}}\rightarrow \mathcal{D}%
^{\otimes 2}\otimes \mathcal{O}(\mathcal{P+F})\rightarrow 0.
\label{subsheaf_04}
\end{equation}

One might alternatively and equivalently denote super Riemann surfaces by
the pair $(\mathcal{S},\mathcal{D})$ and define a generalised spin structure
by $\mathcal{D}^{s}\overset{\mathrm{def}}{=}\mathcal{D}^{\ast }|_{C}$,
following \cite{Witten:2012ga}. Here $\mathcal{D}$ is a choice of
distribution on $\mathcal{S}$, and $\mathcal{D}^{s}$ is a generalised spin
structure on the reduced space $C$. If $\mathcal{S}$ is a $(1|1)$%
-dimensional supermanifold over a Riemann surface $C$, then every
distribution $\mathcal{D}\subset T_{\mathcal{S}}$ satisfying (\ref%
{subsheaf_04}) is in bijective correspondence with generalised spin
structures $\mathcal{D}^{s}~$on $C$.

Consequently, viewing $\mathfrak{M}_{g,n,n^{\prime }}$ as a supermanifold
and using the description of tangent spaces of supermanifolds (modulo the
fermionic ideal), at any isomorphism class $[(\mathcal{S},\mathcal{D})]\in
\mathfrak{M}_{g,n,n^{\prime }}$ we have:
\begin{eqnarray}
T_{\mathfrak{M}_{g,n,n^{\prime }},+}|_{[(\mathcal{S},\mathcal{D})]}
&=&H^{1}(C,T_{C}\otimes \mathcal{O}(-\mathcal{P}-\mathcal{F})),~\ \ ~~~~~
\notag \\
~T_{\mathfrak{M}_{g,n,n^{\prime }},-}|_{[(\mathcal{S},\mathcal{D})]}
&=&H^{1}(C,\mathcal{D}^{s\ast })  \label{tangent bundle}
\end{eqnarray}%
where $\mathcal{D}^{s}$ is the spin structure on $C$ which is uniquely
associated to the distribution $\mathcal{D}$ and $\mathcal{D}^{s\ast }$ is
its dual.

In (\ref{berezinian 01}) we have a general expression for the Berezinian of
a supermanifold. When $\mathfrak{X}=\mathfrak{M}_{g,n,n^{\prime }}$ is a
supermoduli space, its tangent bundle is described in (\ref{tangent bundle}%
). Then by Serre duality we find, at a super Riemann surface class $[(%
\mathcal{S},\mathcal{D})]$ with underlying Riemann surface $C$, its
cotangent space is given by:
\begin{equation}
T_{\mathfrak{M}_{g,n,n^{\prime }},+}^{\ast }|_{[(\mathcal{S},\mathcal{D}%
)]}=H^{0}(C,\omega _{C}^{\otimes 2}\otimes \mathcal{O}(\mathcal{P+F}%
)),~~~~~~~~T_{\mathfrak{M}_{g,n,n^{\prime }},-}^{\ast }|_{[(\mathcal{S},%
\mathcal{D})]}=H^{0}(C,\omega _{C}\otimes \mathcal{D}^{s})
\end{equation}%
where $\omega _{C}$ is the canonical bundle on the curve $C$. Using (\ref%
{berezinian 01}) now, the fiber of the Berezinian of $\mathfrak{M}%
_{g,n,n^{\prime }}$ at a point $[(\mathcal{S},\mathcal{D})]$ on supermoduli
space is:
\begin{equation}
\mathrm{Ber}~\mathfrak{M}_{g,n,n^{\prime }}|_{[(\mathcal{S},\mathcal{D}%
)]}\cong \frac{H^{0}(C,\omega _{C}^{\otimes 2}\otimes \mathcal{O}(\mathcal{P}%
+\mathcal{F}))}{H^{0}(C,\omega _{C}\otimes \mathcal{D}^{s})},
\label{berezinian_06}
\end{equation}%
modulo the fermionic ideal in $\mathcal{O}_{\mathfrak{M}_{g,n,n^{\prime }}}$.

The classical Mumford isomorphism between certain line bundles over the
moduli space of Riemann surfaces can be generalised to isomorphisms between
certain Berezinians over the supermoduli space $\mathfrak{M}_{g}$. With $%
L_{g}$ a line bundle with fiber $L_{g}|_{[C]}=\det H^{0}(C,\omega _{C})$,
the bosonic Mumford isomorphism is $L_{g}^{~n}\cong L_{g}^{\otimes
(6n^{2}-6n+1)}$. Over super Riemann surfaces we can define a line bundle or,
more generally, a sheaf $L_{g}^{3/2}$ with fiber $L_{g}^{3/2}|_{[(\mathcal{S}%
,\mathcal{D})}=\det H^{0}(C,\omega _{C}\otimes \mathcal{D}^{s})$. In
supergeometry, the Berezinian plays the role of the determinants and so
taking Berezinians and tensor powers leads to the following.

\begin{lemm}
Over a genus $g$ super Riemann surface class $[(\mathcal{S},\mathcal{D})]$
with underlying Riemann surface $C$, we can define a family of sheaves $%
L_{g}^{~n,m}$ with fiber
\begin{equation}
L_{g}^{~n,m}|_{[(\mathcal{S},\mathcal{D})]}=\mathrm{Ber}\left\{
H^{0}(C,\omega _{C}^{\otimes n})\oplus H^{0}(C,(\omega _{C}\otimes \mathcal{D%
}^{s})^{\otimes m})\right\} .  \label{sheaves_04}
\end{equation}%
Then%
\begin{equation}
L_{g}^{~n,m}\cong L_{g}^{~n}\otimes \left( L_{g}^{\frac{3m}{2}}\right)
^{\ast }.  \label{relation_04}
\end{equation}%
\label{family_sheaves}
\end{lemm}

\textit{Proof. }Since $(\mathcal{D}^{s})^{\otimes 2}\cong \omega _{C}$, we
can set $\mathcal{D}^{s}=\omega _{C}^{1/2}$ and thereby identify $%
H^{0}(C,(\omega _{C}\otimes \mathcal{D}^{s})^{\otimes m})$ with $L_{g}^{~%
\frac{3m}{2}}$. Hence we have that $L_{g}^{~n,m}\cong L_{g}^{~n}\otimes
(L_{g}^{3m/2})^{\ast }$. {\hfill $\square $} \newline

We see that (\ref{berezinian_06}) is a particular case of (\ref{sheaves_04})
for $(n,m)=(2,1)$, and hence $\mathrm{Ber}~\mathfrak{M}_{g}=L_{g}^{~2,1}$.
The Mumford isomorphism realises the line bundles $L_{g}^{~n}$ as a certain
tensor powers of $L_{g}=L_{g}^{~1}$. With $3m/2=m^{\prime }+1/2$ in (\ref%
{relation_04}), for $m^{\prime }$ the integral part, we obtain from the
classical Mumford isomorphism,
\begin{align}
L_{g}^{~n,m}& \cong L_{g}^{~n}\otimes \left( L_{g}^{m^{\prime }+\frac{1}{2}%
}\right) ^{\ast }  \notag \\
& \cong L_{g}^{\otimes (6n^{2}-6n+1)}\otimes \left( L_{g}^{\otimes
(6m^{\prime 2}-1/2)}\right) ^{\ast }.  \label{isomorphism 05}
\end{align}%
For $(n,m)=(2,1)$, from (\ref{relation_04}) and (\ref{isomorphism 05}) above
we find,
\begin{equation}
\mathrm{Ber}~\mathfrak{M}_{g}\cong L_{g}^{~\otimes 13}\otimes \left(
L_{g}^{\otimes \frac{11}{2}}\right) ^{\ast }\cong (L_{g}^{~3/2})^{5}.
\label{isomorphism 07}
\end{equation}%
This is the Mumford isomorphism for the Berezinian of supermoduli space,
i.e., the super Mumford isomorphism. Now in (\ref{isomorphism 07}) we see
that the right-hand side is a tensor product of powers of the line bundle $%
L_{g}=L_{g}^{~1}$ and $(L_{g})^{\ast }=L_{g}^{~-1}$.

There is a variant of this description. In \cite{Witten:2013tpa} and \cite%
{Voronov:1988ms}, the authors use the Mumford isomorphism over supermoduli
space in a slightly different form to that in (\ref{isomorphism 07}). Note
from (\ref{tangent bundle}) that%
\begin{equation}
\left( \mathrm{Ber}~T_{\mathfrak{M}_{g}}\right) ^{\ast }\cong
L_{g}^{~2}\otimes \left( L_{g}^{~\frac{1}{2}}\right) ^{\ast }\cong
L_{g}^{~3/2}.
\end{equation}%
Hence we find%
\begin{equation}
\mathrm{Ber}~\mathfrak{M}_{g}~\cong ~(L_{g}^{~3/2})^{5}~\cong (\left(
\mathrm{Ber}~T_{\mathfrak{M}_{g}}\right) ^{\ast })^{5}=\left( \mathrm{Ber}%
~T_{\mathfrak{M}_{g}}\right) ^{-5}.  \label{isomorphism 09}
\end{equation}%
The isomorphism $\mathrm{Ber}~\mathfrak{M}_{g}\cong \left( \mathrm{Ber}~T_{%
\mathfrak{M}_{g}}\right) ^{-5}$ in (\ref{isomorphism 09}) is used by Witten
\cite{Witten:2013tpa} for superstring perturbation theory. We have seen that
this also follows from our Lemma \ref{family_sheaves}.

To understand the relation to the super Mumford form, firstly observe that
the super Mumford isomorphism in %Upon realising that the
%The super Mumford \emph{form} can be constructed from the super Mumford isomorphism upon realising that
(\ref{isomorphism 09}) is equivalent to requiring $\mathrm{Ber}$ $T_{%
\overline{\mathfrak{M}}_{g}}^{\ast }\otimes (\mathrm{Ber}$ $T_{\overline{%
\mathfrak{M}}_{g}})^{5}$ be holomorphically trivial, meaning $\mathrm{Ber}$ $%
T_{\overline{\mathfrak{M}}_{g}}^{\ast }\otimes (\mathrm{Ber}$ $T_{\overline{%
\mathfrak{M}}_{g}})^{5}\cong \mathcal{O}_{\overline{\mathfrak{M}}_{g}}$. The
genus $g$ super Mumford form, denoted $\Psi _{g}$, is then a global section
of this trivial bundle, i.e., that $\Psi _{g}\in \Gamma (\overline{\mathfrak{%
M}}_{g},\mathrm{Ber}$ $T_{\overline{\mathfrak{M}}_{g}}^{\ast }\otimes (%
\mathrm{Ber}$ $T_{\overline{\mathfrak{M}}_{g}})^{5})$. These isomorphisms
and forms can be generalised to the case where the super Riemann surfaces
have $n$ Neveu-Schwarz and $n^{\prime }$ Ramond punctures, leading to super
Mumford forms $\Psi _{g,n,n^{\prime }}$ over $\overline{\mathfrak{M}}%
_{g,n,n^{\prime }}$.

\vspace{1pt}

\vspace{1pt}

\vspace{1pt}

\subsection{Boundary contributions and partial compactifications of
supermoduli spaces}

The vacuum superstring amplitude at $g$ loop order is an integral of the
superstring measure, which ought to be computed not over $\mathfrak{M}_{g}$
but its compactification $\overline{\mathfrak{M}}_{g}$. The total, $g$ loop
superstring amplitude therefore receives contributions from the bulk $%
\mathfrak{M}_{g}$ and from the boundary $\partial \overline{\mathfrak{M}}%
_{g} $. Witten \cite{Witten:2015hwa} looks at the boundary in genus $g=2$
and observes that contributions to D'Hoker and Phong's derivation of the
superstring amplitude to two-loop order will be vanishing. This is
consistent with the vanishing of the two-loop vacuum amplitude obtained by
D'Hoker and Phong. In this paper we look at boundary contributions to the
three-loop and higher loop amplitudes. As discussed in previous sections,
the boundary of supermoduli space has codimension $(1|0)$ and parametrises
super Riemann surfaces of lower genus. In particular while $\overline{%
\mathfrak{M}}_{g,n,n^{\prime }}$, for a given $(g,n,n^{\prime })$ may have
complications related to projectability of its bulk $\mathfrak{M}%
_{g,n,n^{\prime }}$, some of its lower genus boundary components may
nevertheless fiber over their reduced space with odd dimensional fibers or
holomorphically map to a bosonic space. Now for dimensional reasons, $%
\mathfrak{M}_{g}$ is split for genus $g=0,1$, and hence projected, i.e., can
be holomorphically fibered over its reduced space. In contrast, Donagi and
Witten \cite{Donagi:2013dua} showed that: $\mathfrak{M}_{g,n,n^{\prime }}$
(and so also its compactification $\overline{\mathfrak{M}}_{g,n,n^{\prime }}$%
) will be \emph{non}-projected for
\begin{equation}
g-1\geq n+n^{\prime }\geq 1.  \label{range_03}
\end{equation}%
Note in particular that $\mathfrak{M}_{2,1}$ will be non-projected.
Surprisingly however, $\mathfrak{M}_{2;+}$ in fact \emph{is} projected as
ilustrated by D'Hoker and Phong in their computation of the superstring
vacuum amplitude at two loop order.

The projection $\mathfrak{M}_{2;+}\rightarrow \mathcal{S}\mathcal{M}%
_{2;+}\rightarrow \mathcal{M}_{2}$ constructed by D'Hoker and Phong uses the
identification of $g=2$ super Riemann surfaces $\mathcal{S}$ with their
period matrices $\Omega (\mathcal{S})$; and a formula relating the period
matrix $\Omega (\mathcal{S})$ to the period matrix of the underlying Riemann
surface $C$, denoted $\Omega (C)$. In a particular gauge, termed \emph{split
gauge}, this formula identifies $\Omega (\mathcal{S})$ with $\Omega (C)$,
thereby leading to the holomorphic projection $\mathcal{S}\mapsto \Omega (%
\mathcal{S})\equiv \Omega (C)\mapsto C$. Since $\mathfrak{M}_{2,1}$ and, by
extension, $\overline{\mathfrak{M}}_{2,1}$ are non-projected by (\ref%
{range_03}), an analogous procedure to that performed by D'Hoker and Phong
will not result in a holomorphic projection. It results instead in a
meromorphic projection $\mathfrak{M}_{2,1;+}\rightarrow \mathcal{S}\mathcal{M%
}_{2,1;+}$. It is singular along the divisor parametrising NS degenerations.
However, in composing the projection $\pi _{2;+}:\mathfrak{M}%
_{2;+}\rightarrow \mathcal{S}\mathcal{M}_{2;+}$ with the forgetful map $%
\mathfrak{M}_{2,1;+}\overset{p}{\rightarrow }\mathfrak{M}_{2;+}$, we obtain
a holomorphic map%
\begin{equation}
q_{2,1;+}:\mathfrak{M}_{2,1;+}\overset{p}{\longrightarrow }\mathfrak{M}_{2;+}%
\overset{\pi _{2}}{\longrightarrow }\mathcal{S}\mathcal{M}_{2;+}.
\label{morphism_q-21}
\end{equation}%
In this way, measures on $\mathfrak{M}_{2,1;+}$ can be reduced to measures
on $\mathcal{S}\mathcal{M}_{2;+}$ and, upon summing over the even spin
structures (GSO projection), to measures on $\mathcal{M}_{2}$. Now more
generally, if $\mu $ is an integration measure on $\mathfrak{M}_{2,1;+}$,
then along $q_{2,1;+}$ we have by the pushforward formula
\begin{equation}
q_{2,1;+\ast }((q_{2,1;+}^{\ast }f)\mu )=f~q_{2,1;+\ast }\mu
\end{equation}%
for any function $f$ on $\mathcal{S}\mathcal{M}_{2}$. Extending to the
compactification leads to an integration relation:
\begin{equation}
\int_{\overline{\mathfrak{M}}_{2,1};+}(q_{2,1;+}^{\ast }f)\mu =\int_{%
\overline{\mathcal{S}\mathcal{M}_{2}};+}f~q_{2,1;+\ast }\mu
\label{integral 04}
\end{equation}%
where now $f$ must be compactly supported. Note that since the fiber of $%
q_{2,1;+}$ is $(1|1)$-dimensional, i.e., not purely odd, the pushforward $%
q_{2,1;+\ast }$ will \emph{not} coincide with the familiar
Berezin-integration. Indeed, Berezin integration will reduce measures on $%
\mathfrak{M}_{2,1;+}$ to $\mathcal{S}\mathcal{M}_{2,1;+}$ along the
meromorphic projection $\pi _{2,1;+}:\mathfrak{M}_{2,1;+}\rightarrow
\mathcal{S}\mathcal{M}_{2,1;+}$. As a result, the Berezin integration of $%
\mu $ along $\pi _{2,1}$ will introduce singularities in the pushed-forward
measure $\pi _{2,1;+\ast }\mu $.

The integration formula in (\ref{integral 04}) involves the
compactifications of supermoduli space and spin moduli space. Witten in \cite%
[Sec. 5]{Witten:2013tpa}, observes however that the D'Hoker-Phong projection
$\pi _{2;+}:\mathfrak{M}_{2;+}\rightarrow \mathcal{S}\mathcal{M}_{2;+}$ does
not extend to a holomorphic projection of the compactification $\overline{%
\mathfrak{M}}_{2;+}$. Indeed, along the divisor $\mathcal{D}_{NS;sep.;-}$
parametrising separating NS degenerations of type $(-,-)$, $\pi _{2;+}$
fails to both (1) be holomorphic along $\mathcal{D}_{NS;sep.;-}$; and (2) to
project $\mathcal{D}_{NS;sep.;-}$ onto its reduced space. As a result, the
extension to the boundary is generally meromorphic, so there exists a
commutative diagram:
\begin{equation}
\xymatrix{ \Mfr_{2;+} \ar[d]_{\pi_2} \ar[r]^\subset &
\overline{\Mfr}_{2;+}\ar[d]^{\overline{\pi}_{2;}} \\ \Scl\Mcl_{2;+}
\ar[r]^\subset & \overline{\Scl\Mcl}_{2;+} }  \label{diagram 04}
\end{equation}%
where $\pi _{2;+}$ is the holomorphic D'Hoker-Phong projection, and $%
\overline{\pi }_{2;+}$ is a meromorphic extension to the compactification.
There are however boundary components along which $\pi _{2}$ will extend
holomorphically. One such boundary component is the divisor $\mathcal{D}%
_{NS;sep.;+}$ parametrising separating NS degenerations of type $(+,+)$. It
was noted by Witten in \cite[Sec. 5]{Witten:2013tpa}, that $\overline{\pi }%
_{2;+}$ will both (1) be holomorphic along $\mathcal{D}_{NS;sep.;+}$; and
(2) project $\mathcal{D}_{NS;sep.;+}$ onto its reduced space. The relation
in (\ref{integral 04}) will therefore be valid near the divisor $\mathcal{D}%
_{NS;sep.;+}$.

Now recall that the vacuum amplitude at $g$-loop order is obtained by
integration over $\overline{\mathfrak{M}}_{g}$. A \emph{partial} $g$-loop
boundary contribution is then a contribution from the factor $\overline{%
\mathfrak{M}}_{2,1}$ in the boundary. As explained above however, the
holomorphic, D'Hoker-Phong projection $\pi _{2;+}:\mathfrak{M}%
_{2;+}\rightarrow \mathcal{S}\mathcal{M}_{2;+}$ will extend generally to a
meromorphic projection upon full compactification; and as noted in the
comments succeeding (\ref{diagram 04}), $\overline{\pi }_{2;+}$ will be
holomorphic along the NS separating divisor $\mathcal{D}_{NS;sep.;+}$. Hence
we can look at the `partial' compactification%
\begin{equation}
\overline{\mathfrak{M}}_{g;NS;sep.}:=\mathfrak{M}_{g}\cup \mathcal{D}%
_{NS;sep.}\subset \overline{\mathfrak{M}}_{g},
\end{equation}%
where super Riemann surfaces are only allowed to develop NS-nodes. For the
even part we have,
\begin{equation}
\overline{\mathfrak{M}}_{2;NS;sep.;+}:=\mathfrak{M}_{2;+}\cup \mathcal{D}%
_{NS;sep.;+}\subset \overline{\mathfrak{M}}_{2;+}.
\label{compactification 03}
\end{equation}%
The projection map $\overline{\mathfrak{M}}_{2;NS;sep.;+}\overset{\overline{%
\pi }_{2;+}}{\longrightarrow }\overline{\mathcal{S}\mathcal{M}}_{2;+}$, is
holomorphic.

More generally now, for any superspace $Y$ with integration measure $\mu
_{Y} $ and holomorphic map%
\begin{equation}
\rho :Y\rightarrow \overline{\mathfrak{M}}_{2;+},  \label{superspace general}
\end{equation}%
we can integrate $\rho _{\ast }\mu _{Y}$ along the partial compactification $%
\overline{\mathfrak{M}}_{2;NS;sep.;+}$ as in (\ref{integral 04}). The
integral of superstring measures along this partial compactification will be
referred to as a partial $g$-loop contribution by using the genus two
supermoduli space as a building block, or factor.

The boundary of $\overline{\mathfrak{M}}_{g}$, denoted $\partial \overline{%
\mathfrak{M}}_{g}$, has codimension $(1|0)$. This means the normal bundle $%
\nu _{g}$ to the embedding $\partial \overline{\mathfrak{M}}_{g}\subset
\overline{\mathfrak{M}}_{g}$ has rank $(1|0)$ and can locally be
parametrised by one \emph{even} variable. The conormal bundle sequence to
this embedding is
\begin{equation}
0\longrightarrow \nu _{g}^{\ast }\longrightarrow \Omega _{\overline{%
\mathfrak{M}}_{g}}^{1}|_{\partial \overline{\mathfrak{M}}_{g}}%
\longrightarrow \Omega _{\partial \overline{\mathfrak{M}}_{g}}^{1}%
\longrightarrow 0.  \label{conormal 03}
\end{equation}%
Taking the Berezinian of (\ref{conormal 03}) and using that $\nu _{g}^{\ast
} $ has rank-$(1|0)$ therefore gives the isomorphism
\begin{equation}
\mathrm{Ber}~\overline{\mathfrak{M}}_{g}|_{\partial \overline{\mathfrak{M}}%
_{g}}\cong \nu _{g}^{\ast }\otimes \mathrm{Ber}~\partial \overline{\mathfrak{%
M}}_{g}.  \label{isomorphism 06}
\end{equation}%
The isomorphism (\ref{isomorphism 06}) relates measures on $\overline{%
\mathfrak{M}}_{g}$ with measures on the boundary $\partial \overline{%
\mathfrak{M}}_{g}$. With the characterisation of the boundary components in (%
\ref{boundary_03}) and (\ref{boundary_04}), the isomorphism in (\ref%
{isomorphism 06}) leads to ansatz factorisations for the superstring measure
near specified boundary components, e.g., $\mathrm{Ber}~\partial \overline{%
\mathfrak{M}}_{g}|_{\Delta _{g_{1},g-g_{1}}}\cong \mathrm{Ber~}\overline{%
\mathfrak{M}}_{g_{1},1}\otimes \mathrm{Ber~}\overline{\mathfrak{M}}%
_{g-g_{1},1}$.

The configuration space for the superstring vacuum states at $g$-loop order
is $\overline{\mathfrak{M}}_{g}$. Along a NS separating divisor $\mathcal{D}%
_{NS;sep.}\subset \partial \overline{\mathfrak{M}}_{g}$ with generic
component in (\ref{boundary_03}), observe that the number of punctures of
each factor will always satisfy the inequality (\ref{range_03}) when $g\geq
2 $. Hence $\mathcal{D}_{NS;sep.}$ cannot be projected for any $g\geq 2$. As
in the case of stable curves however, the forgetful map from punctured
supermoduli space $\overline{\mathfrak{M}}_{g,1}\overset{p}{\rightarrow }%
\overline{\mathfrak{M}}_{g}$ realises $\overline{\mathfrak{M}}_{g,1}$ as the
universal family of genus $g$, super Riemann surfaces over $\overline{%
\mathfrak{M}}_{g}$. And so, forgetting the puncture yields a morphism $%
\mathcal{D}_{NS,sep.}\cong \overline{\mathfrak{M}}_{g_{1},1}\times \overline{%
\mathfrak{M}}_{g_{2},1}\rightarrow \overline{\mathfrak{M}}_{g_{1}}\times
\overline{\mathfrak{M}}_{g_{2}}$, where $g_{1}+g_{2}=g$. If either of the
factors $\overline{\mathfrak{M}}_{g_{1}}$ or $\overline{\mathfrak{M}}%
_{g_{2}} $ are projected, then measures on $\mathcal{D}_{NS,sep.}$ can be
pushed-forward and integrated over the projected factor $\overline{\mathfrak{%
M}}_{g_{1}}$ or $\overline{\mathfrak{M}}_{g_{2}}$. In this way, components
of the integration measure over $\mathcal{D}_{NS,sep.}$ can be calculated by
integrating along the composite morphism
\begin{equation}
\mathcal{D}_{NS,sep.}\cong \overline{\mathfrak{M}}_{g_{1},1}\times \overline{%
\mathfrak{M}}_{g_{2},1}\rightarrow \overline{\mathfrak{M}}_{g_{1}}~~~\text{%
\textrm{or~~~}}\mathcal{D}_{NS,sep.}\cong \overline{\mathfrak{M}}%
_{g_{1},1}\times \overline{\mathfrak{M}}_{g_{2},1}\rightarrow \overline{%
\mathfrak{M}}_{g_{2}}.
\end{equation}%
Now recall the morphism $q_{2,1;+}$ from (\ref{morphism_q-21}). Extending it
to the compactification results in a map,
\begin{equation}
\bar{q}_{2,1;+}:\overline{\mathfrak{M}}_{2,1;+}\overset{p}{\longrightarrow }%
\overline{\mathfrak{M}}_{2;+}\overset{\overline{\pi }_{2;+}}{\longrightarrow
}\overline{\mathcal{SM}}_{2;+}.  \label{holomorphic}
\end{equation}%
Specialising $\overline{q}_{2,1;+}$ to the partial compactification along $%
(+,+)$-separating NS nodes in (\ref{compactification 03}) then gives a
holomorphic map. In genus $g\geq 2$ now, setting $g_{1}=2$ so that $%
g_{2}=g-2 $ gives the boundary component:
\begin{equation}
\begin{array}{ccc}
\overline{\mathfrak{M}}_{2,1;+}\times \overline{\mathfrak{M}}_{g-2,1;\pm } &
\overset{\alpha }{\xrightarrow{\hspace*{0.5cm}}} & \overline{\mathfrak{M}}%
_{g;\pm } \\
{\Big\downarrow}%
\begin{array}{c}
_{{p}} \\
\\
\end{array}
&  &  \\
\overline{\mathfrak{M}}_{2;+} &  & \quad \quad \\
\quad {\Big\downarrow}%
\begin{array}{c}
_{\overline{\pi }_{2;+}} \\
\\
\end{array}
&  &  \\
\overline{\mathcal{SM}}_{2;+} &  &
\end{array}
\label{morphism 02}
\end{equation}%
Note that the above diagram is compatible with both even and odd components
of $\overline{\mathfrak{M}}_{g}$ and $\overline{\mathfrak{M}}_{g-2,1}$
respectively. In this way we obtain a holomorphic map $\mathcal{D}%
_{NS;sep.}\rightarrow \overline{\mathfrak{M}}_{2;+}$. This is the special
case of $Y=\mathcal{D}_{NS;sep.}$ in (\ref{superspace general}). In genus $%
g\geq 2$ then, we can obtain a partial, $g$-loop vacuum contribution from
the boundary component $\mathcal{D}_{NS;sep.}$. In the case of a
non-separating, NS degeneration $\mathcal{D}_{NS;nsep.}\subset \partial
\overline{\mathfrak{M}}_{g}$, recall from (\ref{boundary_03}) that $\mathcal{%
D}_{NS;nsep.}\cong \overline{\mathfrak{M}}_{g-1,2}$. The inequality in (\ref%
{range_03}) holds for genus $g\geq 4$; while in genus $g=3$, (\ref{range_03}%
) does not hold. Nevertheless, in genus $g=3$, projecting out the two
NS-punctures yields:
\begin{equation}
\mathcal{D}_{NS;nsep.;+}\overset{\cong }{\longrightarrow }\overline{%
\mathfrak{M}}_{2,2;+}\overset{p^{2}}{\longrightarrow }\overline{\mathfrak{M}}%
_{2;+}.
\end{equation}%
As in the general case in (\ref{morphism 02}) then, we can obtain a partial,
three-loop boundary contribution from the non-separating divisor $\mathcal{D}%
_{NS;nsep.;+}$ to the three-loop superstring vacuum amplitude.

Recall from (\ref{boundary_04}) that the divisors parametrising
degenerations along Ramond punctures are fibered over supermoduli spaces
with $(0|1)$-dimensional fiber. The key difference here between calculating
partial three loop contributions from these boundary components lies
therefore in firstly integrating out the odd fiber parameter along the given
fibrations $\pi _{R;sep.;+}$ and $\pi _{R;nsep.;+}$ in (\ref{boundary_04})
respectively. This allows for reducing the integration measure to measures
on supermoduli spaces. An analysis similar to the case of NS-degenerations
can be undertaken to evaluate partial three-loop contributions along the
divisors parametrising degenerations along Ramond punctures.

The below diagram is another example for a higher genus, namely $g=4$:
\begin{equation}
\begin{array}{ccc}
\overline{\mathfrak{M}}_{2,1;+}\times \overline{\mathfrak{M}}_{2,1;+} &
\overset{\alpha }{\xrightarrow{\hspace*{0.5cm}}} & \overline{\mathfrak{M}}%
_{4;+} \\
\quad \quad {\Big\downarrow}%
\begin{array}{c}
_{{p}_{2}\times {p}_{2}} \\
\\
\end{array}
&  &  \\
\overline{\mathfrak{M}}_{2;+}\times \overline{\mathfrak{M}}_{2;+} &  & \quad
\quad \\
\quad \quad ~\quad {\Big\downarrow}%
\begin{array}{c}
_{\overline{\pi }_{2;+}\times \overline{\pi }_{2;+}} \\
\\
\end{array}
&  &  \\
\overline{\mathcal{SM}}_{2;+}\times \overline{\mathcal{SM}}_{2;+} &  &
\end{array}
\label{diagram 05}
\end{equation}%
We will illustrate the example of $g=3$ in Section 3.5. In some special
cases in the context of $g=3,4$, both factors could be projectable as in
above diagram (\ref{diagram 05}).

\vspace{1pt}

\subsection{Degenerations and super Mumford forms}

\label{degeneration-super-Mumford}

In analogy with the Mumford relations on the moduli space of Riemann
surfaces, there are analogous relations between divisors on the supermoduli
space, as illustrated in (\ref{isomorphism 09}). Generalising these
relations to the punctured case leads to a trivialising section of a certain
tensor product involving canonical bundles, $\Psi _{g,n,n^{\prime }}$ on $%
\overline{\mathfrak{M}}_{g,n,n^{\prime }}$, referred to as the super Mumford
form. This global section $\Psi _{g,n,n^{\prime }}$ defines a holomorphic
measure on $\overline{\mathfrak{M}}_{g,n,n^{\prime }}$ and, in perturbative
superstring theory, is taken to be the $g$ loop, superstring measure with $%
n+n^{\prime }$ external states. Its integral over $\overline{\mathfrak{M}}%
_{g,n,n^{\prime }}$ gives the $g$ loop scattering amplitude. There are a
number of issues surrounding the computation of this amplitude however,
arising primarily from the complexity in the geometry of the supermoduli
space itself.

\vspace{1pt}

For instance, $\Psi _{g,0}$ corresponds to vacuum amplitudes where there are
no external particles; and $\Psi _{g,1}$ corresponds to tadpole amplitudes
or one-point amplitudes. The form $\Psi _{g,2}$ corresponds to two-point
amplitudes or propagators of a string state. The tadpole graphs are those
with only one external state or external particle. The tadpole amplitudes
describe the amplitudes for a particle to disappear into the vacuum. In
unitary superstring theory, the tadpole amplitude for a massless stable
particle is zero.

\vspace{1pt}\vspace{1pt}

An attractive feature of looking at the boundary of supermoduli space is
that, as is clear from (\ref{boundary_03}), (\ref{boundary_04}), its
components parametrise super Riemann surfaces of lower genera. Furthermore,
near the boundary, the superstring measure itself factorises into lower
genus components, c.f., the discussions around (\ref{isomorphism 06}). In
genus $g=g_{1}+g_{2}$, we have the following $\alpha $- and $\beta $-type
clutchings such as
\begin{equation}
\overline{\mathfrak{M}}_{g_{1},n_{1}+1,n_{1}^{\prime }}\times \overline{%
\mathfrak{M}}_{g_{2,},n_{2}+1,n_{2}^{\prime }}\overset{\alpha }{%
\longrightarrow }\overline{\mathfrak{M}}_{g,n,n^{\prime }},~~~~~~\ ~~%
\overline{\mathfrak{M}}_{g-1,n+2,n^{\prime }}\overset{\beta }{%
\longrightarrow }\overline{\mathfrak{M}}_{g,n,n^{\prime }}.
\end{equation}%
The images of $\alpha $ and $\beta $ form components of the boundary $%
\partial \overline{\mathfrak{M}}_{g,n,n^{\prime }}$ which we denote $\Delta
_{\alpha }$ and $\Delta _{\beta }$ respectively. We use the geometric model
of gluing maps to analyse boundary contributions, via observations by Witten
\cite{Witten:2012bh,Witten:2013tpa}. Along a separating or non-separating
degeneration respectively, with $g=g_{1}+g_{2}$, the super Mumford form
satisfies the ansatz factorisation
\begin{equation}
\Psi _{g,n,n^{\prime }}|_{\Delta _{\alpha }}\sim \Psi
_{g_{1},n_{1}+1,n_{1}^{\prime }}F(\varepsilon )\Psi
_{g_{2},n_{2}+1,n_{2}^{\prime }},~~~~~~~~\Psi _{g,n,n^{\prime }}|_{\Delta
_{\beta }}\sim \Psi _{g-1,n+2,n^{\prime }}G(\varepsilon ),
\label{asymptotics 01}
\end{equation}%
for some possible singular forms $F(\varepsilon )$ and $G(\varepsilon )$
depending on the degeneration parameter $\varepsilon $. The tilde sign $\sim
$ means up to a multiplicative constant, which is a normalisation constant
in the definition of $\Psi $. By (\ref{asymptotics 01}) then, the super
Mumford form is expected to satisfy relevant asymptotics near the boundary
divisors. Integrating these asymptotic expressions for $\Psi _{g,n,n^{\prime
}}$ along the boundary reveals how the $g$-loop amplitude will receive
contributions from the boundary of $\overline{\mathfrak{M}}_{g,n,n^{\prime
}} $.

\vspace{1pt}

Away from the boundary divisor, the node is smoothed out into a thin
cylinder. The closer to the boundary divisor in the supermoduli space, the
thinner the cylinder is. The thin cylinder means a long distance propagation
of a closed string state in string theory. From the worldsheet
superconformal field theory point of view, gluing along the NS punctures as
in (\ref{gluing 01}) is equivalent to insertion of the form$~V_{\ell
}(a|\alpha )\otimes V_{r}(b|\beta )\,\varepsilon ^{m}\mathrm{d}\varepsilon $%
, where $\varepsilon \,$\ is the gluing parameter and $V_{\ell }$ and $V_{r}$
are operators inserted on the two sides of the cylinder at points $a|\alpha $
and $b|\beta $. The conformal dimension of the operator is $1+\frac{m}{2}$
for $m\geq -2.$

The forms $F(\varepsilon )$ and $G(\varepsilon )$ are defined on the
cylinder described by the degeneration parameter $\varepsilon $. A form with
a general order $m$ is given by the expression
\begin{equation}
\lbrack \mathrm{d}a|\mathrm{d}\alpha ]\varepsilon ^{m}\mathrm{d}\varepsilon
\lbrack \mathrm{d}b|\mathrm{d}\beta ].
\end{equation}%
In the gluing relations (\ref{gluing 01}), notice that $\varepsilon $ is
related to the square-root of the NS gluing parameter $q_{NS}$. To
investigate this relation further, we can consider a more general change of
variable from $\varepsilon $ to a new gluing parameter $q$ as follows,
\begin{equation}
(-q)^{1/2}=\varepsilon +\varepsilon ^{p}C\alpha \beta
\end{equation}%
or equivalently $q=-\varepsilon (\varepsilon +2C\varepsilon ^{p}\alpha \beta
)$ with $p$ a general real number. Integrating over the odd moduli $\alpha $
and $\beta $ while fixing $q$ then yields,
\begin{eqnarray}
\int [\mathrm{d}a|\mathrm{d}\alpha ]\varepsilon ^{m}\mathrm{d}\varepsilon
\lbrack \mathrm{d}b|\mathrm{d}\beta ] &\sim &\int [\mathrm{d}\alpha ]((%
\mathrm{d}aCq^{\frac{1}{2}(m+p-2)}\mathrm{d}q\mathrm{d}b)\alpha \beta +\dots
)[\mathrm{d}\beta ]  \notag \\
&\sim &\int ~\mathrm{d}a~Cq^{\frac{1}{2}(m+p-2)}\mathrm{d}q~\mathrm{d}b~.
\label{integral 02}
\end{eqnarray}%
Here $\mathrm{d}a$, $\mathrm{d}q$, $\mathrm{d}b$ are independent integration
variables. Accordingly, near a boundary component $\Delta _{j}$ the super
Mumford form factorises as follows,
\begin{equation}
\Psi _{g,n,n^{\prime }}|_{\Delta _{j}}\sim \Psi _{\Delta _{j}}\frac{\mathrm{d%
}q}{q^{\frac{1}{2}(2-m-p)}}  \label{boundary 06}
\end{equation}%
where $\Psi _{\Delta _{j}}\in \Gamma (\mathrm{Ber}$ $T_{\Delta _{j}}^{\ast
}\otimes (\mathrm{Ber~}T_{\Delta _{j}})^{5})$\ and $\Psi _{g,n,n^{\prime
}}\in \Gamma (\mathrm{Ber}$ $T_{\overline{\mathfrak{M}}_{g,n,n^{\prime
}}}^{\ast }\otimes (\mathrm{Ber}$ $T_{\overline{\mathfrak{M}}_{g,n,n^{\prime
}}})^{5})$ has a pole of a certain order on the boundary $\Delta _{j}\subset
\overline{\mathfrak{M}}_{g,n,n^{\prime }}$. This formula (\ref{boundary 06})
is derived from the gluing map of the geometric model by analytic methods,
and is also a realisation of the factorisation in (\ref{isomorphism 06}).
The integration on the reduced space of the cylinder is
\begin{equation}
\int \frac{\mathrm{d}^{2}q}{q^{\frac{1}{2}(2-m-p)}{\bar{q}}^{\frac{1}{2}%
(2-m-p)}}
\end{equation}%
which would have a pole for $m+p<2$. This physically means a propagator of a
closed string state on a cylinder. In fact, the Neveu-Schwarz (or Ramond)
node is the limit when the circle winding the cylinder, for a propagating
Neveu-Schwarz state (or Ramond state), is shrinking to zero size.

In the case where the boundary divisor parametrises separating NS
degenerations of $(+,+)$-type, Witten \cite{Witten:2012bh,Witten:2012ga}
derived,
\begin{equation}
\Psi _{g;\pm }\sim \Psi _{g_{1};+}[\mathrm{d}a|\mathrm{d}\alpha ]\frac{%
\mathrm{d}\varepsilon }{\varepsilon ^{2}}[\mathrm{d}b|\mathrm{d}\beta ]\Psi
_{g_{2};\pm }.
\end{equation}%
This corresponds to $m=-2,p=0$. Note that there is always a $\Psi _{g_{1};+}$
factor with $+$ spin structure in the above factorisation. At a $(-,-)$
degeneration \cite{Witten:2012bh,Witten:2012ga},
\begin{equation}
\Psi _{g;+}\sim \Psi _{g_{1};-}(V)\,[\mathrm{d}a|\mathrm{d}\alpha ]\mathrm{d}%
\varepsilon \,\varepsilon ^{8}[\mathrm{d}b|\mathrm{d}\beta ]\,\Psi
_{g_{2;}-}(V),  \label{degeneration_07}
\end{equation}%
where $\Psi _{g_{i};-}(V)$ is computed by inserting the superconformal
primary operator $V$ of dimension 5, and $m=8,p=-2$. Note that this term
scales as a positive power of $\varepsilon \,$.$~$After change of variable,
this term scales as a positive power of $q$. Hence the integration of this
term (\ref{degeneration_07}) is vanishing in the limit of $\varepsilon
\rightarrow 0~$or $q\rightarrow 0$, and so the contribution to the $g$-loop
amplitude from this boundary component is vanishing.

For non-separating degeneration of NS type we have the factorisation,
\begin{equation}
\Psi _{g}\sim \Psi _{g-1}[\mathrm{d}a|\mathrm{d}\alpha ]\frac{\mathrm{d}%
\varepsilon }{\varepsilon ^{2}}[\mathrm{d}b|\mathrm{d}\beta ].
\end{equation}%
Changing variable from $\varepsilon$ to $q$ via $(-q)^{1/2}=\varepsilon
+\alpha \beta ~$\cite{Witten:2012bh} and integrating over $\alpha$ and $%
\beta $ with fixed $q$ gives,
\begin{equation}
\int [\mathrm{d}\alpha ]\varepsilon ^{-2}\mathrm{d}\varepsilon \lbrack
\mathrm{d}\beta ]\sim \int [\mathrm{d}\alpha ](q^{-2}\mathrm{d}q~\alpha
\beta +\dots )[\mathrm{d}\beta ]=\int q^{-2}\mathrm{d}q,
\label{change of variable 02}
\end{equation}%
and hence:
\begin{eqnarray}
\pi _{\ast }\Psi _{g}|_{\mathrm{im}~\alpha _{g_{1},g_{2}}} &\sim &\Psi
_{g_{1}}^{\mathrm{d}a}\frac{\mathrm{d}q}{q^{2}}\Psi _{g_{2}}^{\mathrm{d}b}.
\label{degeneration_09} \\
\pi _{\ast }\Psi _{g}|_{\mathrm{im}~\beta _{g-1}} &\sim &\Psi _{g-1}^{%
\mathrm{d}a~\mathrm{d}b}~\frac{\mathrm{d}q}{q^{2}}.
\end{eqnarray}%
Here $\pi _{\ast }$ denotes integrating out two fermionic moduli e.g. $%
\mathrm{d}\alpha \mathrm{d}\beta $. Furthermore, we have used forgetful
morphsims in defining the forms appearing on the right hand side above,
e.g., $\Psi _{g_{1},1}=p^{\ast }(\Psi _{g_{1}})$. The form $\Psi _{g_{1}}^{%
\mathrm{d}a}$ is then the contraction of $\Psi _{g_{1}}$ along $\mathrm{d}a$%
. Eq. (\ref{degeneration_09}), like (\ref{integral 02}), contains factors
like $\mathrm{d}a\mathrm{d}q\mathrm{d}b$ with independent variables for
integration.

From the worldsheet superconformal field theory point of view \cite%
{Witten:2012bh,Witten:2012ga}, gluing along the Ramond punctures as in (\ref%
{gluing 02}) is equivalent to insertions of the form
\begin{equation}
e^{\zeta G_{0}}[\mathrm{d}\zeta ]\frac{\mathrm{d}q_{R}}{q_{R}},
\end{equation}%
where $q_{R}$ and $\zeta $ are bosonic and fermionic gluing parameters and $%
G_{0}$ is the zero-mode of the worldsheet supercurrent. The factor $e^{\zeta
G_{0}}$ is due to the coupling of the worldsheet gravitino with the
worldsheet supercurrent in the action. The $\mathrm{d}\zeta $ is integrated
over $\mathbb{C}^{0|1}$. From the point of view of the boundary divisor, the
$\mathbb{C}^{0|1}$ is the extra fermionic fiber of the boundary divisor with
Ramond degeneration in the supermoduli space.

For non-separating degeneration of Ramond type,
\begin{equation}
\Psi _{g}|_{\mathrm{im}~\beta _{g-1}}\sim \sum_{\alpha _{i},\alpha _{j}}\Psi
_{g-1,2^{\prime }}(\Xi _{\alpha _{i}},\Xi _{\alpha _{j}})\frac{\mathrm{d}%
q_{R}}{q_{R}}e^{\zeta G_{0}}[\mathrm{d}\zeta ].  \label{ramond_02}
\end{equation}%
Here $\Psi _{g-1;2^{\prime }}(\Xi _{\alpha _{1}},\Xi _{\alpha _{2}})$ is the
super Mumford form for a genus $g-1$ super Riemann surfaces with $n^{\prime
}=2$ Ramond punctures and with the superconformal primary operators $\Xi
_{\alpha _{1}},\Xi _{\alpha _{2}}$ inserted at these punctures \cite%
{Witten:2013tpa}. Recall that in the notation we use primes to denote the
Ramond puncture number.

For separating degenerations of Ramond type near the boundary of the
supermoduli space,
\begin{equation}
\Psi _{g,2^{\prime }}(\Xi _{\alpha _{1}},\Xi _{\alpha _{2}})|_{\mathrm{im}%
~\alpha _{g_{1},g_{2}}}\sim \sum_{\alpha _{i},\alpha _{j}}\Psi
_{g_{1},2^{\prime }}(\Xi _{\alpha _{1}},\Xi _{\alpha _{i}})\frac{\mathrm{d}%
q_{R}}{q_{R}}e^{\zeta G_{0}}[\mathrm{d}\zeta ]\Psi _{g_{2},2^{\prime }}(\Xi
_{\alpha _{2}},\Xi _{\alpha _{j}}),  \label{sep_06}
\end{equation}%
where $g_{1}+g_{2}=g$.$~$Note that (\ref{sep_06}) is for the case when there
are two external massless string states on the left-hand side of (\ref%
{sep_06}).

The factorisation is valid near the degeneration, when $|\varepsilon |$ or $%
|q_{R}|$ are small respectively, and the degeneration is the limit when $%
\varepsilon \rightarrow 0$ or $q_{R}\rightarrow 0$ respectively.

\subsection{Integration in supergeometry and supermoduli spaces}

To calculate the superstring amplitude, integration on the supermoduli space
is needed. This would involve firstly integrating out the odd (fermionic)
coordinates, and then integrating the resulting measure over a classical
moduli space. In this section, we first describe integration of holomorphic
measures. We then describe integration over smooth supermanifolds and over
complex supermanifolds. Then we present a useful integration formula (\ref%
{integral 05}) which will be used in Section 3.5. Finally, we describe
integral forms which are also relevant to the question of integrating along
the boundary of the supermoduli space.

On smooth, real, orientable manifolds $M$ there is, up to a positive
constant, a natural and unique volume measure $\nu _{M}$. This $\nu _{M}$ is
a global section of the line bundle of volume forms
\begin{equation}
\det T^{\ast }M=\Omega ^{\dim M}(M),  \label{volume 02}
\end{equation}%
where $\Omega ^{1}(M)=\Gamma (M,T^{\ast }M)$ are the global sections of the
cotangent vector bundle over $M$. Now we describe integration of holomorphic
measures. \noindent Complex manifolds $X$ are smooth manifolds equipped with
a choice of integrable complex structure. Any $n$-dimensional complex
manifold $X$ will have an underlying $2n$-dimensional \emph{real} manifold
which we denote by $X^{\infty }$. With $J$ an integrable, complex structure
we can identify $X=(X^{\infty },J)$. Volume forms on $X$ can be integrated
over $M=X^{\infty }$ as a smooth real manifold. We want to describe \emph{%
holomorphic} volume measures however which come from the complex manifold $X$%
. This requires understanding the decomposition of the differential forms on
$X^{\infty }$. Let $TX^{\infty \ast }$ denote the cotangent bundle. With
respect to the complex structure $J$ we have a decomposition into
holomorphic and anti-holomorphic forms $TX^{\infty \ast }\cong
T^{1,0}X^{\infty \ast }\oplus T^{0,1}X^{\infty \ast }$ and hence a tensor
product factorisation of the volume forms on $X^{\infty }$. Now with $%
T^{\ast }X$ the holomorphic cotangent bundle of $X$ there is a natural
inclusion $T^{\ast }X\subset T^{1,0}X^{\infty \ast }$; and similarly an
inclusion of the anti-holomorphic cotangent bundle $\overline{T^{\ast }X}%
\subset T^{0,1}X^{\infty \ast }$. The inclusions are at the level of the
sections of vector bundles. Denote $\det X=\det T^{\ast }X$ and $\det
\overline{X}=\det \overline{T^{\ast }X}$. Using (\ref{volume 02}) we have
therefore an inclusion into the volume forms on $X^{\infty }$,
\begin{equation}
\det X\otimes \det \overline{X}\subset \det T^{1,0}X^{\infty \ast }\otimes
\det T^{0,1}X^{\infty \ast }\cong \det X^{\infty }.  \label{volume 03}
\end{equation}%
Sections of $\det X$ are referred to as holomorphic volume forms on $X$; and
similarly sections of $\det \overline{X}$ are anti-holomorphic volume forms.
Complex conjugation $z\mapsto \overline{z}$ induces a conjugation on
holomorphic forms. In particular, to any holomorphic volume form $\omega \in
\det X$ we have the conjugate-squaring%
\begin{equation}
\det X\longrightarrow \det X\otimes \det \overline{X},\ \ \ \ \ \ \ \ \omega
\longmapsto \omega \otimes \overline{\omega }.  \label{volume 04}
\end{equation}%
Composing (\ref{volume 03}) with (\ref{volume 04}) gives a mapping between
volume forms $\det X\rightarrow \det X^{\infty }$. And so, with this
mapping, we can integrate holomorphic functions against holomorphic volume
forms on $X$ by simply integrating the resulting $(n,n)$-form over the
underlying real manifold $X^{\infty }$. Explicitly, for a holomorphic volume
form $\omega $ on a complex manifold $X$, we have
\begin{equation}
\int_{X}\omega \overset{\mathrm{def}}{=}\int_{X^{\infty }}\omega \otimes
\overline{\omega }.  \label{volume 05}
\end{equation}

Before turning to integration in supergeometry, we discuss a notion in
algebraic topology serving to motivate subsequent notions in supergeometry.
A vector bundle $E$ over a manifold $M$ is fibered over $M$ with linear
fibers. With $\pi :E\rightarrow M$ denoting the fibration, compactly
supported forms on $E$ can be formally reduced to forms on $M$ via the
integration-along-fiber map $\pi _{\ast }:\Omega _{cpct.}^{j}(E)\rightarrow
\Omega ^{j-\mathrm{rank}~E}(M)$. As explained by Bott and Tu \cite{Bott Tu},
if $x$ denote local coordinates on the base $M$ and $y$ coordinates on the
fiber, then $(x,y)$ will be local coordinates on $E$ and $\pi _{\ast }$ is
defined on $\Omega _{cpct.}^{j}(E)$ by:
\begin{equation}
\pi _{\ast }:(\pi ^{\ast }f)\mathrm{d}x_{1}\cdots \mathrm{d}x_{m}\mathrm{d}%
y_{1}\cdots \mathrm{d}y_{m^{\prime }}\longmapsto \left\{
\begin{array}{ll}
f(x)~\mathrm{d}x_{1}\cdots \mathrm{d}x_{j-\mathrm{rank}~E} &
\mbox{if $m^\prime =
\mathrm{rank}~E$} \\
0 & \mbox{otherwise}.%
\end{array}%
\right.
\end{equation}%
Intuitively, that $\pi _{\ast }(\mathrm{d}y_{1}\cdots \mathrm{d}y_{m^{\prime
}})=1$ if $m^{\prime }=\mathrm{rank}~E$ and is zero otherwise. Berezin \cite%
{Berezin} defined integration over supermanifolds analogously to the
integration-along-fiber map above. Crucially, this definition only makes
sense if the supermanifold can be fibered over its reduced space with purely
odd fibers---a property known as `projected'. That any supermanifold can be
smoothly projected over its reduced space is a consequence of Batchelor's
splitting theorem \cite{Batchelor}. Holomorphically however, there are
generally obstructions to fibering (or, projecting) complex supermanifolds
over their reduced spaces. Hence, holomorphic measures over complex
supermanifolds cannot generally be integrated in the way outlined by
Berezin. In the case where complex supermanifolds $\mathfrak{X}$ are \emph{%
non-projected}, i.e., cannot be holomorphically fibered over their reduced
space, it is an open question as to how to integrate holomorphic measures
over $\mathfrak{X}$. Donagi and Witten \cite{Donagi:2013dua,Donagi:2014hza}
found that $\mathfrak{M}_{g}$ cannot be globally holomorphically projected
onto its reduced space for any genus greater or equal to five.

To continue our discussion of the integration on supermanifolds now, there
are two kinds of objects which can be integrated over supermanifolds. They
are (1) Berezinian volume forms; and (2) integral forms. The former are
similar to volume forms on manifolds as discussed above; and the latter are
similar to distributions. We will firstly consider Berezinian volume forms.

There is no unique `top form' on a supermanifold since the differential of
odd, or Grassmann, variables are no longer nilpotent. For example, if $%
\theta $ is odd, then $\mathrm{d}\theta \wedge \mathrm{d}\theta \neq 0$. One
can nevertheless form the module of volume forms on a supermanifold
analogous to the determinant line bundle from (\ref{volume 02}). To any
super vector space $\mathbb{V}=V\oplus \Pi W$, where $\Pi W$ is the vector
space of Grassmann variables with fermionic statistics, we can form its
Berezinian $\mathrm{Ber}~\mathbb{V}$, which is a $(1|1)$-dimensional vector
space. Now for a supermanifold $\mathfrak{X}$ its cotangent bundle $T^{\ast }%
\mathfrak{X}$ is is a bundle of super vector spaces. It makes sense to
therefore set $\mathrm{Ber}~\mathfrak{X}=\mathrm{Ber}~T^{\ast }\mathfrak{X}$%
. To see how to integrate these volume forms, let $|\mathfrak{X}|$ be the
reduced space of $\mathfrak{X}$. It is smooth manifold and embeds naturally
inside $\mathfrak{X}$. With a projection map $\pi :\mathfrak{X}\rightarrow |%
\mathfrak{X}|$, the supermanifold $\mathfrak{X}$ can be realised as fibered
over its reduced space with odd or `fermionic fiber'. Generalising the
classical integration-along-fiber construction in differential topology, we
can use $\pi $ to integrate out the fermionic fibers to recover thereby a
volume form on $|\mathfrak{X}|$. Denoting by $\pi _{*}$ the
integration-along-fiber map, we have therefore a morphism of sheaves $\pi
_{\ast }:\mathrm{Ber}~\mathfrak{X}\rightarrow \det |\mathfrak{X}|$. With $%
\pi $ then we can define, for any $\sigma _{\mathfrak{X}}\in \Gamma (%
\mathfrak{X},\mathrm{Ber}~\mathfrak{X})$:
\begin{equation}
\int_{\mathfrak{X}}\sigma _{\mathfrak{X}}\overset{\mathrm{def}}{=}\int_{|%
\mathfrak{X}|}\pi _{\ast }\sigma _{\mathfrak{X}}.  \label{integral 03}
\end{equation}%
Now suppose $\mathfrak{X}$ is endowed with a covering $\mathfrak{U}=(%
\mathfrak{U}_{\alpha })$ where each $\mathfrak{U}_{\alpha }$ is isomorphic
to $(|\mathfrak{U}_{\alpha }|,C^{\infty }(|\mathfrak{U}_{\alpha }|)\otimes
\wedge ^{\bullet }\mathbb{R}^{q})$ where $|\mathfrak{U}_{\alpha }|$ is the
reduced space of $\mathfrak{U}_{\alpha }$. Denote the odd dimension of $%
\mathfrak{X}$ by $q$. Let $F\in \mathcal{O}_{\mathfrak{X}}(\mathfrak{X})$ be
a global, smooth function. Then over $\mathfrak{U}_{\alpha }$ we have $F|_{%
\mathfrak{U}_{\alpha }}\in \mathcal{O}_{\mathfrak{X}}(\mathfrak{U}_{\alpha
})\cong C^{\infty }(|\mathfrak{U}_{\alpha }|)\otimes \wedge ^{\bullet }%
\mathbb{R}^{q}$. With the projection map $\pi :\mathfrak{X}\rightarrow |%
\mathfrak{X}|$, we can write $F|_{\mathfrak{U}_{\alpha }}=(\pi ^{\ast
}g_{\alpha })\otimes \Theta _{\alpha }$ for some $g_{\alpha }\in C^{\infty
}(|\mathfrak{U}_{\alpha }|)$ and a Grassmann constant $\Theta _{\alpha }\in
\wedge ^{\bullet }\mathbb{R}^{q}$. Note that this constant can be absorbed
into the Berezinian volume form $\sigma _{\mathfrak{X},\alpha }=\sigma _{%
\mathfrak{X}}|_{\mathfrak{U}_{\alpha }}$. Finally now, in order to ensure
the ultimate integral is well defined, choose a partition of unity $\rho _{|%
\mathfrak{U}|}$ subordinate to $|\mathfrak{U}|$ and set $f_{\alpha }=\rho
_{\alpha }g_{\alpha }$. For each index $\alpha $, $\rho _{\alpha }$ is
compactly supported in $|\mathfrak{U}_{\alpha }|$. Then over $\mathfrak{U}%
_{\alpha }$ we have by (\ref{integral 03}),
\begin{equation}
\int_{\mathfrak{U}_{\alpha }}F|_{\mathfrak{U}_{\alpha }}\sigma _{\alpha
}=\int_{|\mathfrak{X}|}f_{\alpha }~\pi _{\ast }(\Theta _{\alpha }\sigma _{%
\mathfrak{X},\alpha }).
\end{equation}%
We can integrate against any volume form $\sigma \in \Gamma (\mathfrak{X},%
\mathrm{Ber}~\mathfrak{X})$,
\begin{equation}
\int_{\mathfrak{X}}F\sigma _{\mathfrak{X}}=\sum_{\alpha }\int_{\mathfrak{U}%
_{\alpha }}F_{\alpha }\sigma _{\mathfrak{X},\alpha }.
\end{equation}%
By (\ref{integral 03}), the integration above only depends on a choice of
projection $\pi :\mathfrak{X}\rightarrow |\mathfrak{X}|$.

We now describe integration over complex supermanifolds. Firstly, there are
a number of ways to define a complex supermanifold. For our purposes, a
complex supermanifold $\mathfrak{Y}$ is a supermanifold where: (1) the
reduced space $|\mathfrak{Y}|$ of $\mathfrak{Y}$ is a complex manifold and;
(2) the restriction of the tangent bundle $T\mathfrak{Y}$ to $|\mathfrak{Y}|$
is holomorphic. The tangent bundle of $\mathfrak{Y}$ is $\mathbb{Z}_{2}$%
-graded, so $T\mathfrak{Y}\cong T_{+}\mathfrak{Y}\oplus T_{-}\mathfrak{Y}$.
Restricting $T\mathfrak{Y}$ to $|\mathfrak{Y}|$ gives $T\mathfrak{Y}|_{|%
\mathfrak{Y}|}\cong T|\mathfrak{Y}|\oplus N|\mathfrak{Y}|$, where $N|%
\mathfrak{Y}|\rightarrow |\mathfrak{Y}|$ is a vector bundle. If $|\mathfrak{Y%
}|$ is a complex manifold, then $T|\mathfrak{Y}|$ will automatically be
holomorphic. Condition (2) then amounts to requiring $N|\mathfrak{Y}|$ also
be holomorphic. As in the case of complex manifolds, for complex
supermanifolds there will be an underlying smooth supermanifold which we
denote by $\mathfrak{Y}^{\infty }$. Our conventions here are such that the
complex structure is only defined by reference to the \emph{even}
coordinates. And so, if $(z|\theta )$ denote local coordinates on $\mathfrak{%
Y}$, their conjugation is $\widetilde{(z|\theta )}=(\widetilde{z}|\theta )$.
As explained by Witten in \cite{Witten:2012bg}, the conjugate $\widetilde{z}$
coincides with the familiar complex conjugate $\overline{z}$ on the reduced
space $|\mathfrak{Y}^{\infty }|$. That is, along the embedding $|\mathfrak{Y}%
^{\infty }|\subset \mathfrak{Y}^{\infty }$ we have $\widetilde{(z|0)}=(%
\overline{z}|0)$. As a consequence of this convention, the conjugate
supermanifold $\widetilde{\mathfrak{Y}}$ only differs from $\mathfrak{Y}$ in
that the reduced space is conjugate, i.e, $|\overline{\mathfrak{Y}}|=%
\overline{|\mathfrak{Y}|}$. Where the odd tangent bundle is concerned
however, we have $T_{-}\widetilde{\mathfrak{Y}}=T_{-}\mathfrak{Y}$.

Now let $\pi :\mathfrak{Y}\rightarrow |\mathfrak{Y}|$ be a holomorphic
projection. On Berezinian volume forms it defines the mapping, $(f+g\theta )[%
\mathrm{d}x|\mathrm{d}\theta ]\overset{\pi _{\ast }}{\longmapsto }g~dx$, for
$x$ denoting a complex, even variable $x$ and $\theta $ the odd variable.
For multiple even variables $x_{1},...,x_{m}$ and odd variables $\theta
_{1},...,\theta _{n}$,%
\begin{equation}
~\theta _{1}\cdots \theta _{n}~[\mathrm{d}x_{1}\cdots \mathrm{d}x_{m}|%
\mathrm{d}\theta _{1}\cdots \mathrm{d}\theta _{n}]\overset{\pi _{\ast }}{%
\longmapsto }\mathrm{d}x_{1}\cdots \mathrm{d}x_{m}.
\end{equation}%
Then as in the smooth case, $\pi $ realises $\mathfrak{Y}$ as being
holomorphically fibered over its reduced space with fermionic fibers. With $%
\mathrm{Ber}~\mathfrak{Y}$ the space of holomorphic volume forms on $%
\mathfrak{Y}$, and $\det |\mathfrak{Y}|$ the holomorphic volume forms on the
reduced space $|\mathfrak{Y}|$, the integration-along-fiber map gives a
relation $\pi _{\ast }:\mathrm{Ber}~\mathfrak{Y}\rightarrow \det |\mathfrak{Y%
}|$. Recall that by our conventions here, we only conjugate the even
parameters. As such, and since the even and odd parameters are globally
distinguished on $\mathfrak{Y}$, there is no need to implement an operation
as in (\ref{volume 04}) for volume forms on $\mathfrak{Y}$ directly. We can
instead defer this operation to the reduced space $|\mathfrak{Y}|$.
Therefore, for a volume form $\Psi \in \Gamma (\mathfrak{Y},\mathrm{Ber}~%
\mathfrak{Y})$ and a holomorphic projection $\pi :\mathfrak{Y}\rightarrow |%
\mathfrak{Y}|$, we can define:
\begin{equation}
\int_{\mathfrak{Y}}\Psi \overset{\mathrm{def}}{=}\int_{|\mathfrak{Y}|}\pi
_{\ast }\Psi \otimes \overline{\pi _{\ast }\Psi }.  \label{integral 05}
\end{equation}%
Note that the right-hand side above does not make any reference to the
underlying, smooth supermanifold $\mathfrak{Y}^{\infty }$. The procedure for
calculating $\pi _{\ast }\Psi $ however is similar to that for volume forms
on smooth supermanifolds $\mathfrak{X}$ since $\pi $ here is holomorphic.
This can be viewed as being in analogy with the integration of holomorphic
volume forms in (\ref{volume 05}). In Section 3.5, we use the definition (%
\ref{integral 05}) extensively in forming our integrations.

The projection map $\pi :\mathfrak{X}\rightarrow |\mathfrak{X}|$ fibering a
smooth supermanifold over its reduced space always exists, albeit
non-canonically so. This is in contrast to the complex case where
holomorphic projections $\pi :\mathfrak{Y}\rightarrow |\mathfrak{Y}|$ need
not exist. The formula (\ref{integral 05}) is suitable only in the case
where $\mathfrak{Y}$ is such that $\pi $ exists as a holomorphic map. As
mentioned earlier, when $\mathfrak{Y}=\overline{\mathfrak{M}}_{g}$ is the
supermoduli space of genus $g$ curves, Donagi and Witten in \cite%
{Donagi:2013dua} illustrated precisely this: that a holomorphic projection $%
\pi :\overline{\mathfrak{M}}_{g}\rightarrow |\overline{\mathfrak{M}}_{g}| $
does not exist when $g\geq 5$.

Another type of object that can be integrated over supermanifolds are \emph{%
integral forms}. The integration of integral forms on supermanfolds was
discussed in \cite{Bernstein Leites 01,Bernstein Leites 02,Witten:2012bg}. A
particularly appealing feature of the codimension-one integral forms lies in
their relation to a generalised Stokes' Theorem.

Suppose $\mathfrak{X}$ is a supermanifold with boundary $\partial \mathfrak{X%
}$ and de Rham differential $\mathrm{d}$. Stokes' Theorem asserts that any
codimension-one, compactly supported integral form $\nu $ satisfies
\begin{equation}
\int_{\mathfrak{X}}\mathrm{d}\nu =\int_{\partial \mathfrak{X}}\nu .
\label{Stokes 02}
\end{equation}%
Specialising to supermoduli space then, for any codimension-one integral
form $\nu $ on $\overline{\mathfrak{M}}_{g,n,n^{\prime }}$, we have by (\ref%
{Stokes 02}) that $\int_{\overline{\mathfrak{M}}_{g_{1}+g_{2},n_{1}+n_{1}^{%
\prime },n_{2}+n_{2}^{\prime }}}\mathrm{d}\nu =\int_{\partial \overline{%
\mathfrak{M}}_{g_{1}+g_{2},n_{1}+n_{1}^{\prime },n_{2}+n_{2}^{\prime }}}\nu $%
. By (\ref{inclusions}), the images of $\alpha $ and $\beta $ form
components of the boundary $\partial \overline{\mathfrak{M}}_{g,n,n^{\prime
}}$. Therefore,
\begin{equation}
\int_{\overline{\mathfrak{M}}_{g,n,n^{\prime }}}\mathrm{d}\nu
=\int_{\partial \overline{\mathfrak{M}}_{g,n,n^{\prime }}}\nu =\int_{\mathrm{%
im}~\alpha }\nu +\int_{\mathrm{im}~\beta }\nu .
\end{equation}%
In this way we see how codimension-one, integral forms will receive
contributions from boundary divisors. Integral forms can also be useful in
describing other observables on the boundary of the supermoduli space, such
as anomalies.

\subsection{Boundary contribution to three loop vacuum amplitude}

As an illustration of the formalisms in the previous sections, we analyse
the three loop vacuum amplitudes in detail in this section. We use the
factorisation of super Mumford forms $\Psi _{3}$ near the boundary of the
supermoduli space to analyse the contribution to the vacuum amplitude at
genus three, from the boundary of the supermoduli space, and consider the
cases of NS and Ramond nodes at the degeneration. The supermoduli space of
the super Riemann surfaces is denoted by $\overline{\mathfrak{M}}_{g}$,
while the spin moduli space of the Riemann surfaces is denoted by $\overline{%
\mathcal{SM}}_{g}$, and the\ moduli space of Riemann surfaces is denoted by $%
\overline{\mathcal{M}}_{g}$.

In this section, $g=g_{1}+g_{2}=3$. Hence the clutching maps describing the
degenerations are:
\begin{eqnarray}
&&\overline{\mathfrak{M}}_{2,1}\times \overline{\mathfrak{M}}_{1,1}\overset{%
\alpha _{2,1}}{\longrightarrow }\overline{\mathfrak{M}}_{3,0}. \\
&&\overline{\mathfrak{M}}_{3,1}\times \overline{\mathfrak{M}}_{0,1}\overset{%
\alpha _{3,0}}{\longrightarrow }\overline{\mathfrak{M}}_{3,0}. \\
&&\overline{\mathfrak{M}}_{2,2}\overset{\beta _{2}}{\longrightarrow }%
\overline{\mathfrak{M}}_{3,0}.
\end{eqnarray}%
Now recall that in genus $g=2$, D'Hoker and Phong constructed a holomorphic
projection $\pi _{2,+}:\mathfrak{M}_{2,+}\rightarrow \mathcal{SM}_{2,+}$.
This projection can be extended to a meromorphic mapping $\pi _{2,1}:%
\mathfrak{M}_{2,1}\rightarrow \mathcal{SM}_{2,1}$ which will be meromorphic
on the compactification. To retain holomorphy however we can, as in (\ref%
{holomorphic}), specialise the morphism $\bar{q}_{2,1;+}:\overline{\mathfrak{%
M}}_{2,1;+}\overset{p}{\longrightarrow }\overline{\mathfrak{M}}_{2;+}\overset%
{\overline{\pi }_{2;+}}{\longrightarrow }\overline{\mathcal{SM}}_{2;+}$ to
the partial compactification formed by allowing $(+,+)$ NS nodes. With the
clutching maps above, we obtain a diagram%
\begin{equation}
\begin{array}{ccc}
\overline{\mathfrak{M}}_{2,1;+}\times \overline{\mathfrak{M}}_{1,1} &
\overset{\alpha }{\xrightarrow{\hspace*{0.5cm}}} & \overline{\mathfrak{M}}%
_{3} \\
\quad {\Big\downarrow}%
\begin{array}{c}
_{{p}_{2}} \\
\\
\end{array}
&  &  \\
\overline{\mathfrak{M}}_{2;+}\times \overline{\mathfrak{M}}_{1,1} &  & \quad
\quad \\
\quad \quad \quad ~~~~{\Big\downarrow}%
\begin{array}{c}
_{\overline{\pi }_{2;+}~\times ~\overline{\pi }_{1,1}} \\
\\
\end{array}
&  &  \\
\overline{\mathcal{SM}}_{2;+}\times \overline{\mathcal{SM}}_{1,1} &  &
\end{array}
\label{diagram}
\end{equation}%
The D'Hoker-Phong projection $\pi _{2,+}:\mathfrak{M}_{2,+}\rightarrow
\mathcal{S}\mathcal{M}_{2,+}$ is defined by sending a genus $g=2$ super
Riemann surface with prescribed period matrix to a genus $g=2$ Riemann
surface with the same period matrix. This mapping generalises to define
mappings $\bar{q}_{2,1;+}\times \overline{\pi }_{1,1}$ in the diagram above
and a projection $\overline{\pi }_{3}:~\overline{\mathfrak{M}}%
_{3}\rightarrow \overline{\mathcal{SM}}_{3}$. However, we do not consider $%
\overline{\pi }_{3}$ here since it is meromorphic. As described in Section %
\ref{degeneration-super-Mumford}, near the boundary of supermoduli space the
super Mumford form admits a factorisation as follows:
\begin{eqnarray}
\Psi _{3}|_{\mathrm{im}~\alpha _{2,1}} &\sim &\Psi _{2,1}F(\varepsilon )\Psi
_{1,1}. \\
\Psi _{3}|_{\mathrm{im}~\alpha _{3,0}} &\sim &\Psi _{3,1}F(\varepsilon )\Psi
_{0,1}. \\
\Psi _{3}|_{\mathrm{im}~\beta _{2}} &\sim &\Psi _{2,2}G(\varepsilon ).
\end{eqnarray}%
Here, the $\varepsilon $ is the degeneration parameter near the boundary,
and $\Psi _{2,1}=p^{\ast }(\Psi _{2})$ for $p$ the forgetful morphism. Note
that while $\overline{\pi }_{3\ast }(\Psi _{3})$ will be singular, $%
\overline{\pi }_{1,1\ast }(\Psi _{1,1})$ will not be singular and $\overline{%
\pi }_{2;+\ast }(\Psi _{2})$ will be non-singular along the $(+,+)$
separating NS divisor. As a result, factors such as $\overline{q}_{2,1;+\ast
}(\Psi _{2,1})~$can be calculated via the D'Hoker-Phong method.

In terms of the gluing parameters $\varepsilon $ and local coordinates near
the punctures $a|\alpha $ and $b|\beta $, we have more explicitly:
\begin{eqnarray}
\Psi _{3}|_{\mathrm{im}~\alpha _{2,1}} &\sim &\Psi _{2}[\mathrm{d}a|\mathrm{d%
}\alpha ]\frac{\mathrm{d}\varepsilon }{\varepsilon ^{2}}[\mathrm{d}b|\mathrm{%
d}\beta ]\Psi _{1}. \\
\Psi _{3}|_{\mathrm{im}~\alpha _{3,0}} &\sim &\Psi _{3}[\mathrm{d}a|\mathrm{d%
}\alpha ]\frac{\mathrm{d}\varepsilon }{\varepsilon ^{2}}[\mathrm{d}b|\mathrm{%
d}\beta ]\Psi _{0}. \\
\Psi _{3}|_{\mathrm{im}~\beta _{2}} &\sim &\Psi _{2}[\mathrm{d}a|\mathrm{d}%
\alpha ]\frac{\mathrm{d}\varepsilon }{\varepsilon ^{2}}[\mathrm{d}b|\mathrm{d%
}\beta ].
\end{eqnarray}%
Changing variable from $\varepsilon $ to $q$ as (\ref{change of variable 02}%
) and integrating over $\alpha $ and $\beta $ with fixed $q$ then gives:
\begin{eqnarray}
\Psi _{3}|_{\mathrm{im}~\alpha _{2,1}} &\sim &\Psi _{2,1}~\frac{\mathrm{d}q}{%
q^{2}}~\Psi _{1,1}.  \label{smf 01} \\
\Psi _{3}|_{\mathrm{im}~\alpha _{3,0}} &\sim &\Psi _{3,1}~\frac{\mathrm{d}q}{%
q^{2}}~\Psi _{0,1}.  \label{smf 02} \\
\Psi _{3}|_{\mathrm{im}~\beta _{2}} &\sim &\Psi _{2,2}~\frac{\mathrm{d}q}{%
q^{2}}.  \label{smf 03}
\end{eqnarray}%
Note, we have used forgetful morphisms in defining the above forms, e.g., $%
\Psi _{2,1}=p^{\ast }\Psi _{2}$.

For non-separating degeneration of Ramond type,
\begin{equation}
\Psi _{3}|_{\mathrm{im}~\beta _{2}}\sim \sum_{\alpha _{1,}\alpha _{2}}\Psi
_{2,2^{\prime }}(\Xi _{\alpha _{1}},\Xi _{\alpha _{2}})e^{\zeta G_{0}}[%
\mathrm{d}\zeta ]\frac{\mathrm{d}q_{R}}{q_{R}}
\end{equation}%
where $\Psi _{2,2^{\prime }}(\Xi _{\alpha _{1}},\Xi _{\alpha _{2}})$ is the
super Mumford form for a genus $2$ super Riemann surface with $n^{\prime }=2$
Ramond punctures and with superconformal primary operators $\Xi _{\alpha
_{1}},\Xi _{\alpha _{2}}$ inserted at these punctures. The terms $q_{R}$ and
$\zeta $ are bosonic and fermionic gluing parameters.

The bosonic gluing parameters $\varepsilon $ or $q_{R}$ can be viewed as the
bosonic coordinate of the fiber of the normal bundle of the boundary
component, in the separating or non-separating cases respectively, c.f., (%
\ref{normal bundle 03}).

In the case of an odd spin structure, the superstring vacuum amplitude is
zero, since one needs the insertion of operators to absorb the ten fermionic
zero-modes of the RNS fermions and for the vacuum amplitude there is no such
operator insertions on the genus three surface. Hence for the three loop
vacuum amplitude, it suffices to consider genus three surfaces with even
spin structures.

The contribution to the superstring amplitude from the boundary of the
supermoduli space is the integration as in (\ref{amplitude_boundary}),
\begin{equation}
\mathcal{A}:=\int_{\partial \overline{\mathfrak{M}}_{g}}\int_{\mathfrak{N}%
}F_{g}.
\end{equation}%
We shall denote by $N$ the reduced space of the fiber $\mathfrak{N~}$of the
normal bundle to the boundary divisor.

For general $g_{1}$ and $g_{2}$ with $g_{1}+g_{2}=g=3$ then, we can use the
factorisation of the super Mumford form from (\ref{smf 01})--(\ref{smf 03})
to get,
\begin{equation}
\mathcal{A}=\int_{\overline{\mathfrak{M}}_{g_{1},1}\times \overline{%
\mathfrak{M}}_{g_{2},1}}\int_{\mathfrak{N}}\Psi _{g_{1},1}\frac{\mathrm{d}q}{%
q^{2}}\Psi _{g_{2},1}.  \label{integral 08}
\end{equation}%
It is known for $g\leq 2$, massless tadpole graphs, i.e. one-point
functions, all vanish in type II superstring theory with unbroken spacetime
supersymmetry \cite{Green:1981yb,DHoker:2001kkt,Witten:2012bh}. That is,
e.g.,
\begin{equation}
\int_{\overline{\mathfrak{M}}_{g_{1},1}}\Psi _{g_{1},1}=0.
\label{tadpole_vanishing_01}
\end{equation}%
Therefore, since the factor in (\ref{integral 08}) contains a tadpole graph,
the amplitude, i.e. the full integral (\ref{integral 08}) will vanish. Now
in the expression (\ref{integral 08}), recall that $\Psi _{g_{i},1}=p^{\ast
}(\Psi _{g_{i}})$, where $p:\overline{\mathfrak{M}}_{g_{i,1}}\rightarrow
\overline{\mathfrak{M}}_{g_{i}}$ is the forgetful morphism. Pushing the
measure forward under $p$ then gives $\int_{\overline{\mathfrak{M}}%
_{g_{i},1}}p^{\ast }(\Psi _{g_{i}})=\int_{\overline{\mathfrak{M}}%
_{g_{i}}}\Psi _{g_{i}}$, c.f., (\ref{integral 04}). The amplitude in (\ref%
{integral 08}) for separating degenerations therefore reduces to,
\begin{equation}
\mathcal{A}=\int_{\overline{\mathfrak{M}}_{g_{1}}\times \overline{\mathfrak{M%
}}_{g_{2}}}\int_{\mathfrak{N}}\Psi _{g_{1}}\frac{\mathrm{d}q}{q^{2}}\Psi
_{g_{2}}.
\end{equation}%
For non-separating degenerations of the NS type we have,
\begin{equation}
\mathcal{A}=\int_{\overline{\mathfrak{M}}_{g-1}}\int_{\mathfrak{N}}\Psi
_{g-1}\frac{\mathrm{d}q}{q^{2}},
\end{equation}%
where we have used$~\int_{\overline{\mathfrak{M}}_{g-1,2}}\Psi
_{g-1,2}=\int_{\overline{\mathfrak{M}}_{g-1,2}}p^{\ast }p^{\ast }(\Psi
_{g-1})$.

For non-separating degenerations of Ramond type,%
\begin{equation}
\mathcal{A}=\int_{\overline{\mathfrak{M}}_{g-1,2^{\prime }}}\int_{\mathfrak{N%
}}\int_{\mathbb{C}^{0|1}}\sum_{\alpha _{1},\alpha _{2}}\Psi _{g-1,2^{\prime
}}(\Xi _{\alpha _{1}},\Xi _{\alpha _{2}})e^{\zeta G_{0}}[d\zeta ]\frac{%
\mathrm{d}q_{R}}{q_{R}}.
\end{equation}%
We have the factorisation by the factor $\int_{\mathfrak{N}}\int_{\mathbb{C}%
^{0|1}}e^{\zeta G_{0}}[\mathrm{d}\zeta ]\frac{\mathrm{d}q_{R}}{q_{R}}$. This
factor can be holomorphically projected to its reduced space. The
integration of $q_{R}$ will be as follows. With an infrared regulator $%
\epsilon $, $\int_{N}\frac{\mathrm{d}^{2}q_{R}}{q_{R}{\bar{q}}_{R}}\sim
\int_{\epsilon \leq |q_{R}|}\frac{\mathrm{d}^{2}q_{R}}{q_{R}{\bar{q}}_{R}}%
\sim \ln \epsilon $ could have had a $\ln \epsilon $ infrared divergence,
however, the other prefactor would be vanishing due to summations of
superconformal operators in the presence of unbroken spacetime supersymmetry.

When $g_{1}$, $g-1$, or $g_{2}\leq 2$, the projections $\overline{\pi }%
_{2;+} $,$~\overline{\pi }_{1,1}$ and$~\overline{\pi }_{0,1}$ to the reduced
spaces (which we denote generally by $\pi $) are holomorphic, as we
discussed in Section 3.2. Note that, for instance, $\overline{\pi }_{1,1}~$%
being holomorphic implies that $\overline{\pi }_{1,0}$ is also holomorphic.
Hence we can use an alternative method by holomorphic projection of one of
the factors involving the lower genus supermoduli space and use the
integration formula (\ref{integral 05}), as we describe as follows. Here $N$
is the reduced space of the fiber $\mathfrak{N~}$of the normal bundle of the
boundary divisor. Because one of the components of the Riemann surface is a
tadpole graph, and the factor of the tadpole graph vanishes, we have e.g.,
\begin{equation}
\int_{\overline{\mathcal{M}}_{g_{1}}}\pi _{\ast }\Psi _{g_{1}}\ {\overline{%
\pi _{\ast }\Psi _{g_{1}}}}\int_{N}\frac{\mathrm{d}^{2}q}{q^{2}{\bar{q}}^{2}}%
=0.  \label{vanishing}
\end{equation}%
With an infrared regulator $\epsilon $, $\int_{N}\frac{\mathrm{d}^{2}q}{q^{2}%
{\bar{q}}^{2}}\sim \int_{\epsilon \leq |q|}\frac{\mathrm{d}^{2}q}{q^{2}{\bar{%
q}}^{2}}\sim \frac{1}{\epsilon ^{2}}$ could have had a $\frac{1}{\epsilon
^{2}}$ infrared divergence, but due to the vanishing of the prefactor which
is a vanishing tadpole graph as in Eq. (\ref{tadpole_vanishing_01}), the
full integral (\ref{vanishing}) will be vanishing. For the clutching of the
second type, with NS degeneration,
\begin{equation}
\mathcal{A}=\int_{\overline{\mathcal{M}}_{g-1}}\int_{N}\pi _{\ast }\Psi
_{g-1}\ \overline{\pi _{\ast }\Psi _{g-1}}~\frac{\mathrm{d}^{2}q}{q^{2}{\bar{%
q}}^{2}}.  \label{vanishing_}
\end{equation}%
D'Hoker and Phong have shown that the two-loop two-point function for
massless NS sector vanish, implying that $\int_{\overline{\mathcal{M}}%
_{g-1}}\pi _{\ast }\Psi _{g-1}~\overline{\pi _{\ast }\Psi _{g-1}}=0$. Hence
the full integral (\ref{vanishing_}) is vanishing.

We also mention that in the case $g_{1}=2$, $g_{2}=1$, we can project both
factors. Then by the integration formula (\ref{integral 05}),
\begin{equation}
\mathcal{A}=\int_{\overline{\mathcal{M}}_{g_{1}}\times \overline{\mathcal{M}}%
_{g_{2}}}\int_{N}\pi _{\ast }\Psi _{g_{1}}\ \overline{\pi _{\ast }\Psi
_{g_{1}}}~\frac{\mathrm{d}^{2}q}{q^{2}{\bar{q}}^{2}}\ \pi _{\ast }\Psi
_{g_{2}}~\overline{\pi _{\ast }\Psi _{g_{2}}}~,
\end{equation}%
which also shows the vanishing.

There are two methods of computations above. We could first factorise and
then project on the reduced space of one of the factors. For some graphs, we
could also first factorise and then project both factors. The two methods
are different ways of going along the arrows in the diagram (\ref{diagram})
as we illustrated.

Our analysis shows that the boundary contribution to the three-loop vacuum
amplitude, from the boundary of the supermoduli space will vanish in closed
oriented type II superstring theory with unbroken spacetime supersymmetry.
Furthermore, we know that the vacuum amplitudes at genus zero, one and two
are also vanishing \cite{Green:1981yb,DHoker:2001kkt,Witten:2012bh}, in the
closed oriented type II superstring theory, with unbroken spacetime
supersymmetry. Here, what we have analysed is the boundary contribution to
the superstring amplitude, not the bulk contribution.

There is an analysis of the superstring amplitude at three loop order from
the bulk of the bosonic moduli space using modular forms \cite%
{DHoker:2004qhf,Cacciatori:2008ay,Grushevsky:2008zm,Matone:2010yv,Cacciatori:2008pj}%
. These results are compatible with our analysis. However the approach there
is not derived from supermoduli space or from manifest supersymmetry. It is
therefore not obvious how to directly relate their bosonic ansatz with the
approach here through supermoduli space.

By similar calculations as presented in this section, we can deduce the
following remark.

\begin{remark}
In closed oriented type II superstring theory in spacetime backgrounds with
unbroken supersymmetry, the contribution to massless tadpole graphs, i.e.
one-point functions, from the boundary of the supermoduli space, is
vanishing at three-loop order.
\end{remark}

The idea behind the above remark is, since the Ramond puncture number are
always even, one only need to add one additional NS puncture on one
component of the super Riemann surface with lower genus. Again one uses the
factorisations. One could also use the forgetful morphism associated to that
NS puncture, which would be equivalent to the formalism of integrated NS
vertex operator.

\vspace{1pt}

\section{Discussion}

\label{sec_discussion}

\vspace{1pt}

One of the main goals in this paper is to obtain an understanding of
scattering amplitudes in perturbative superstring theory. We focussed in
particular on contributions to the superstring amplitude from the boundary
of supermoduli space, which in turn can be described by clutching morphisms
on supermoduli spaces of generally lower genera. The physical interpretation
of these clutching morphisms is that they are related to taking an infrared
or large distance limit of superstring amplitudes. Fundamentally, the
superstring amplitudes can be calculated by integrating the superstring
measure, which itself defines a measure on supermoduli space. Following
earlier work in \cite{Voronov:1988ms,Witten:2013tpa}, we discussed how this
measure could be constructed from the generalisation to supermoduli space of
the classical Mumford isomorphisms on the moduli space of Riemann surfaces.

The superstring measure, or super Mumford form, is a measure defined on
supermoduli space. It is holomorphic on the bulk but, in the
compactification, acquires poles near some components of the boundary. We
can understand this pole behaviour by looking at how the super Mumford form
would factorise into forms over lower genera supermoduli spaces along
specified boundary components.

\vspace{1pt}

In genus two, and so at two loop order, D'Hoker and Phong calculated the
superstring amplitude by integrating the superstring measure over the genus
two supermoduli space. This integration involved a projection of supermoduli
space onto its reduced, bosonic space, where more classical integration
methods could then be applied. In genus three, i.e., at three loop order, it
was not clear how to apply D'Hoker and Phong's method of calculation since
it is not known whether there exists a holomorphic projection of the genus
three supermoduli space onto its reduced, bosonic space. We observe however
that since the boundary in genus three parametrises super Riemann surfaces
of lower genera, some of these components may at least admit a holomorphic
map to a bosonic space. Considering only those boundary components which
admit such a map led to the notion of a partial compactification. Following
observations by Witten in \cite{Witten:2013tpa} we note that the genus two
boundary component parametrising $(++)$ Neveu-Schwarz degenerations of a
genus three super Riemann surface admits precisely such a map.

\vspace{1pt}

To state our result more clearly, our analysis shows that the boundary
contribution to the three loop vacuum amplitude, from the boundary of the
supermoduli space, will vanish in closed, oriented, Type II superstring
theory with unbroken spacetime supersymmetry. It also implies that upon
compactification to four spacetime dimensions preserving the supersymmetry,
the boundary contribution to the cosmological constant at three loop level
is also zero in Type II superstring with unbroken spacetime supersymmetry.

\vspace{1pt}

Furthermore, our observations are compatible with the results on superstring
amplitudes at three loop order in \cite%
{DHoker:2004qhf,Cacciatori:2008ay,Grushevsky:2008zm,Matone:2010yv,Cacciatori:2008pj}
which are obtained using ansatz for modular forms on bosonic moduli space.
It is an open and interesting problem as to how to understand the bulk, or
interior, contribution to the three loop vacuum amplitude from the viewpoint
of supermoduli space. Moreover, it would also be desirable to understand the
relation between the approach from supermoduli space and from modular forms
in the calculation of superstring amplitudes.

\vspace{1pt}

In theories with unbroken supersymmetry the vacuum energy is vanishing since
contributions from bosons and fermions cancel each other exactly. It would
also be good to understand the vanishing of boundary contributions from the
point of view of possible nonrenormalisation theorems \cite%
{Martinec:1986wa,Dine:1986vd} at three loop, as well as from the point of
view of, and relations to, other superstring formalisms such as the
Green-Schwarz formalism and the pure spinor formalism.

\vspace{1pt}

The boundary contributions to the superstring amplitude can also be viewed
as correction terms to calculations of superstring amplitudes over the bulk.
Correction terms in genus three may then be relevant for heterotic string
theory. Calculations in heterotic string theory are more subtle however, as
they involve embedded integration cycles in products of moduli spaces $%
\mathfrak{M}_{g;L}\times \mathcal{M}_{g;R}$ parametrising left and right
movers \cite{Witten:2013cia,DHoker:2013sqy}. Boundary contributions, or
correction terms, may also be relevant in the calculation of amplitudes for
open strings in type II string theory, where there is only one factor of the
chiral measure. In the $\alpha ^{\prime }\rightarrow 0$ limit one can
recover scattering amplitudes for ordinary non-abelian gauge field theories.

\vspace{1pt}

More generally and as mentioned earlier, the boundary of the supermoduli
space is useful in computing the infrared or large distance limit of the
superstring amplitudes. In this limit there appear infinitely thin cylinders
forming an internal component of the worldsheet super Riemann surface.
Contributions to amplitudes from the boundary of supermoduli space have also
been previously considered by e.g., \cite%
{Witten:2013cia,DHoker:2013sqy,Witten:2012ga,Witten:2012bh,Sen:2015uoa}

\vspace{1pt}

Fayet-Iliopoulos terms can be generated at one-loop in superstring theory on
heterotic backgrounds with anomalous $U(1)$ symmetry \cite%
{Dine:1987xk,Atick:1987gy,Dine:1987gj}. As a consequence, it generates a
nonzero mass term at one-loop for tree-level massless scalars charged under
the anomalous $U(1)$. It also induces a nonzero dilaton tadpole and nonzero
vacuum energy \cite{DHoker:2013sqy,Witten:2013cia} at two-loop. There is
tree-level spacetime supersymmetry but it is breaking at one-loop. The
aspect of nonzero tadpole amplitudes and associated spacetime supersymmetry
breaking by string loop effects is a fundamental idea in superstring theory.
A similar degeneration of super Riemann surfaces where there are infinitely
thin cylinders can also occur, e.g., as in \cite{Dine:1987gj}, and so are
related to the boundary of supermoduli spaces. As such, these models are
particularly relevant for approaches involving supermoduli space and its
boundary.

\vspace{1pt}

In this paper we have constrained our analysis to Type II superstring
theory. The analysis for heterotic $SO(32)$ string theory is more involved
\cite{Witten:2013cia,DHoker:2013sqy}. In that case, for certain orbifold
backgrounds with tree-level supersymmetry, spacetime supersymmetry is broken
at one-loop and there is non-zero two-loop vacuum amplitude \cite%
{DHoker:2013sqy,Witten:2013cia}. This is mainly due to the breaking of
supersymmetry by string loop effects. One main difference is that in the
type II theory with unbroken spacetime supersymmetry, we have zero tadpole
amplitudes for stable massless particles.

Factorisations of two loop superstring amplitudes also occur in \cite%
{DHoker:2005jhf} in the type II case, which is useful for checking S-duality
of type II superstring theory. The techniques of \cite{DHoker:2005jhf} are
also useful in computing the normalisation factor for the vacuum energy. The
S-duality covariance and factorisation constraints would be a good
consistency check for computing the amplitudes and the effective action in
Type II superstring and supergravity theory.

\vspace{1pt}

\vspace{1pt}

\vspace{1pt}

\section*{Acknowledgments}

The work was supported in part by Yau Mathematical Sciences Center and
Tsinghua University, and by grant TH-533310008 of Tsinghua University (to
H.L.).

%%%%%%%%%%%%%%%%%%%%%%%%%%%%%%%%%%%%%%%%%%%%%%%%%%%%%%%%%%%%%%%%%%%%%%%%%%%%%

\vspace{1pt}

\vspace{1pt}

\vspace{1pt}

\vspace{1pt}

\vspace{1pt}\appendix

\section{Compactification of moduli space and spin moduli space}

\label{appendix_special_case}

\renewcommand{\theequation}{A.\arabic{equation}} \setcounter{equation}{0} %
\renewcommand{\thethm}{A.\arabic{thm}} \setcounter{thm}{0} %
\renewcommand{\theprop}{A.\arabic{prop}} \setcounter{prop}{0}

In this appendix, in order to be self-contained and for the convenience of
the reader, we briefly overview the compactification of the moduli space of
ordinary Riemann surfaces and the compactification of the spin moduli space
of spin Rieman surfaces. Note that we use the term Riemann surface
interchangeably with (complex) curve.

\vspace{1pt}

The moduli space $\mathcal{M}_{g}$ parametrises families of smooth curves $%
\mathcal{X}$ over a base $B$. The non-compactness of $\mathcal{M}_{g}$
follows from the property that such families a non-complete base $B$ cannot
generally be extended to smooth families of curves over the completion $%
\widehat{B}$, even up to finitely many base-changes. Remarkably however, if
one allows families of curves to have at-worst nodal singularities, then any
family of smooth curves $\mathcal{X}\rightarrow B$ can be extended to a
family of stable curves $\widehat{\mathcal{X}}\rightarrow \widehat{B}$ with
at-worst nodal singularities. A stable curve $C$ has a nodal singularity at
a point $p\in C$ if, in any local, affine coordinate system $(x,y)$ at $p$
that $xy=0$ at $p$. The boundary divisor of the compactification $\partial
\overline{\mathcal{M}}_{g}\subset \overline{\mathcal{M}}_{g}$ consists of
degenerations of two distinct kinds: \emph{separating} and \emph{%
non-separating}. If $D_{sep.}\subset \partial \overline{\mathcal{M}}_{g}$
parametrises separating degenerations, then a generic component of $D_{sep.}$
is isomorphic to the product $\overline{\mathcal{M}}_{g_{1},1}\times
\overline{\mathcal{M}}_{g_{2},1}$ where $g_{1}+g_{2}=g$. If $%
D_{nsep.}\subset \partial \overline{\mathcal{M}}_{g}$ parametrises
non-separating degenerations, then $D_{nsep.}\cong \overline{\mathcal{M}}%
_{g-1,2}$. More generally, with $\overline{\mathcal{M}}_{g,n}$ the
Deligne-Mumford compactification of $n$-pointed, stable, genus $g$ curves,
its boundary $\partial \overline{\mathcal{M}}_{g,n}$ will also parametrise
separating and non-separating degenerations. With $D_{sep.}\subset \partial
\overline{\mathcal{M}}_{g,n}$ parametrising the separating degenerations, a
generic component is isomorphic to the product $\overline{\mathcal{M}}%
_{g_{1},n_{1}+1}\times \overline{\mathcal{M}}_{g_{2},n_{2}+1}$ where $%
n_{1}+n_{2}=n$. With $D_{nsep.}\subset \partial \overline{\mathcal{M}}_{g,n}$
parametrising the non-separating degenerations, $D_{nsep.}\cong \overline{%
\mathcal{M}}_{g-1,n+2}$. The morphisms on boundary components induced by the
inclusion $\partial \overline{\mathcal{M}}_{g,n}\subset \overline{\mathcal{M}%
}_{g,n}$,
\begin{equation}
\overline{\mathcal{M}}_{g_{1},n_{1}+1}\times \overline{\mathcal{M}}%
_{g_{2},n_{2}+1}\overset{\alpha }{\longrightarrow }\overline{\mathcal{M}}%
_{g,n}~~\mbox{and}~~\overline{\mathcal{M}}_{g-1,n+2}\overset{\beta }{%
\longrightarrow }\overline{\mathcal{M}}_{g,n}  \label{clutching_bosonic}
\end{equation}%
are referred to as $\alpha $- and $\beta $-\emph{clutchings} respectively.

\vspace{1pt}

A spin curve is a smooth curve $C$ equipped with a choice of spin structure $%
L_{C}$, which is a line bundle over $C$ satisfying $L_{C}^{\otimes 2}\cong
\omega _{C}$ for $\omega _{C}$ the canonical bundle. The notion of stable,
pointed curves $C$ can be generalised to stable pointed spin curves $%
(C,L_{C})$. Here, $L_{C}$ restricts to a spin structure $L_{C^{sm.}}$ on the
smooth locus of the stable curve $C^{sm.}\subset C$, and vanishes along the
nodes of $C$. Blowing up along the nodes $\{p_{i}\}\in C$ results in a
smooth curve $\pi :\widetilde{C}\rightarrow C$. Off the exceptional divisor,
$\widetilde{C}\setminus \pi ^{-1}(\{p_{i}\})$ is isomorphic to the smooth
locus $C^{sm.}$. As such, the spin structure $L_{C^{sm.}}$ on $C^{sm.}$
pulls back to $\widetilde{C}\setminus \pi ^{-1}(\{p_{i}\})$. The exceptional
divisor $E=\pi ^{-1}(\{p_{i}\})\subset \widetilde{C}$ is isomorphic to a
product of $\mathbb{P}_{\mathbb{C}}^{1}$'s and so the spin structure $%
L_{C^{sm.}}$ can be continued along $E$ by gluing in the standard spin
structure $\mathcal{O}_{\mathbb{P}_{\mathbb{C}}^{1}}(-1)$ on $\mathbb{P}_{%
\mathbb{C}}^{1}$. Cornalba's compactification $\overline{\mathcal{S}\mathcal{%
M}}_{g,n}$ is holomorphically fibered over $\overline{\mathcal{M}}_{g,n}$
with boundary components comprising clutchings, in analogy with (\ref%
{clutching_bosonic}). In contrast to $\overline{\mathcal{M}}_{g}$ however,
the spin structures $L_{C}$ are themselves endowed with parity: $L_{C}$ is
\emph{even} or \emph{odd} if $h^{0}(C,L_{C})\equiv 0$ or $%
h^{0}(C,L_{C})\equiv 1~\mathrm{mod}~2$, respectively. The parity of spin
structures are invariant under continuous deformation. This leads to a
decomposition of the spin moduli space, $\mathcal{S}\mathcal{M}_{g}=\mathcal{%
S}\mathcal{M}_{g}^{+}\cup \mathcal{S}\mathcal{M}_{g}^{-}$. In the
compactification of $\overline{\mathcal{S}\mathcal{M}}_{g}$, one needs
therefore to account for the parity of the spin structure in the separating
degenerations at a node. The analogue of the $\alpha $-clutchings in (\ref%
{clutching_bosonic}) making up the boundary components $\partial \overline{%
\mathcal{S}\mathcal{M}}_{g,n}\subset \overline{\mathcal{S}\mathcal{M}}%
_{g,n}^{+}$ are therefore:
\begin{equation}
\overline{\mathcal{S}\mathcal{M}}_{g_{1},n_{1}+1}^{+}\times \overline{%
\mathcal{S}\mathcal{M}}_{g_{1},n_{2}+1}^{+}\overset{\alpha ^{++}}{%
\longrightarrow }\overline{\mathcal{S}\mathcal{M}}_{g,n}^{+}~~\mbox{and}~~%
\overline{\mathcal{S}\mathcal{M}}_{g_{1},n_{1}+1}^{-}\times \overline{%
\mathcal{S}\mathcal{M}}_{g_{1},n_{2}+1}^{-}\overset{\alpha ^{--}}{%
\longrightarrow }\overline{\mathcal{S}\mathcal{M}}_{g,n}^{+}
\label{alpha_spineven}
\end{equation}%
where $g_{1}+g_{2}=g$ and $n_{1}+n_{2}=n$. Similarly, the $\beta $-clutching
is given by:
\begin{equation}
\overline{\mathcal{S}\mathcal{M}}_{g-1,2}^{+}\overset{\beta ^{+}}{%
\longrightarrow }\overline{\mathcal{S}\mathcal{M}}_{g,n}^{+}.
\label{beta_spinodd}
\end{equation}%
Since clutchings are natural in families, both the $\alpha $- and $\beta $%
-clutchings on the spin moduli space and the moduli space respectively are
compatible with the holomorphic fibration $\overline{\mathcal{S}\mathcal{M}}%
_{g}\rightarrow \overline{\mathcal{M}}_{g}$.

\end{document}